\documentclass[final,5p,times,twocolumn,preprint]{elsarticle}
\usepackage{tikz}
\usepackage{helvet}
\usepackage{color}
\usepackage{xcolor}
\usepackage{graphicx} 
\usepackage{subcaption}
\usepackage{hyperref}
\usepackage{enumitem}
\usepackage{multirow}
\usepackage{threeparttable}
\usepackage{tablefootnote}
\usepackage{float}
\usepackage{amssymb, amsmath}

\def\aap{Astronomy and Astrophysics}

\def\apjl{ApJL}

\def\apjs{ApJS}

\definecolor{mydarkgreen}{RGB}{0,140,0}

\journal{Science Bulletin}

\begin{document}
\begin{frontmatter}
\title{Precise measurements of the cosmic ray proton energy spectrum in the ``knee''  region}
\centerline{\author{\large LHAASO Collaboration\footnotemark[1]$^,$\footnotemark[2]}}
\begin{abstract}
We report the high-purity identification of cosmic-ray (CR) protons and a precise measurement of their energy spectrum from 0.15 to 12 PeV using the Large High Altitude Air Shower Observatory (LHAASO). Abundant event statistics, combined with the simultaneous detection of electrons/photons, muons, and Cherenkov light in air showers, enable spectroscopic measurements with statistical and systematic precision comparable to satellite data at lower energies. The proton spectrum shows significant hardening relative to low-energy extrapolations, culminating at 3 PeV, followed by sharp softening. This distinct spectral structure closely aligned with the knee in the all-particle spectrum points to the emergence of a new CR component at PeV energies that might be linked to the dozens of PeVatrons recently discovered by LHAASO, and offers crucial clues to the origin of Galactic cosmic rays.
\end{abstract}

\end{frontmatter}

\footnotetext[1]{Corresponding authors:  youzhiyong@ihep.ac.cn (Z. You), zhangss@ihep.ac.cn (S. Zhang),  yinlq@ihep.ac.cn (L. Yin) paolo.lipari@roma1.infn.it (P. Lipari).}
\footnotetext[2]{Authors are listed at the end of this paper.}

\section{Introduction}

The ``knee'' in the cosmic ray (CR) all-particle energy spectrum has been known for more than half a century~\cite{cosmicray-knee-1959}, referring to a bending with the spectral index changing from $\sim-2.7$ to $<-3.1$ across the energy of a few Peta-electron Volt (\textcolor{black}{$10^{15}$~eV})~\cite{LHAASO-allparticle}. This was interpreted as indicating the energy limit for CR acceleration by astrophysical sources~\cite{Gaisser:2013bla} or the signature of a change in the properties of CR propagation~\cite{Kachelriess-2019}.
The Large High Altitude Air Shower Observatory 
(LHAASO) has recently confirmed the knee structure in the all-particle spectrum 
and \textcolor{black}{measured} a clear change in composition from relatively light to heavier nuclei across the knee~\cite{LHAASO-allparticle}.
Meanwhile, LHAASO has discovered 43 ultra-high-energy gamma-ray sources in the Milky Way with the maximal energy of photons reaching up to 2.5 PeV~\cite{LHAASO-catalogue,LHAASO-nature-12,LHAASO-sygnus}.
These provide clear evidence of existence of the so-called `PeVatrons' in our Galaxy.
The measurement of the spectra for individual species across the knee, especially for the lightest component, the protons, is crucial for establishing the connection between the celestial sources and the fluxes detected at Earth. 
Direct measurement by space-borne detectors over the past two decades  has previously been the only successful approach so far to collecting pure CR proton samples and revealed that the energy spectrum of protons deviates from a single power-law~\cite{O.AdrianiandL.Pacini2023}.
This approach is effectively limited by energies below 0.1 PeV because of the 
limited geometric acceptance of the current space-borne detectors, such as DAMPE~\cite{DAMPE:2019gys}, CALET~\cite{CALET:2022vro} and ISS-CREAM~\cite{Choi:2022aht}.

Measurements with high statistics of the spectrum of CRs at energies well above 0.1 PeV can only be achieved by using ground-based detector arrays to sample the extensively spread out secondary particles in air showers induced by primary CR particles.
Identifying the primary CR species on the event-by-event basis is  a difficult  task because only secondary particles are collected as they fall down to ground. These particles have undergone a large number of scatterings in the cascade processes in the atmosphere. The arrays are usually designed to measure only a few shower parameters. Many experiments use the muon content as an important parameter in primary particle identification, typically with limited coverage, e.g. with less than 2.3\% filling factor as in CASA-MIA~\cite{CASAMIA-nima1994}, GRAPES~\cite{GRAPES-3:2024mhy} and KASCADE~\cite{kaskade-spectrum-00}. IceCube/IceTop has full coverage of muon detector but at very high threshold energies for muons. Alternative unfolding statistical analysis methods have been developed to obtain the abundance of each species.
In this way, KASCADE measured energy spectra for various components, including protons, assuming different hadronic interaction models. The spectra obtained from different interaction models exhibit large differences~\cite{kaskade-spectrum-00,kaskade-spectrum-01,kaskade-spectrum-02}.
IceCube\&IceTop obtained the proton energy spectrum above 3 PeV by combining the proportions of various components through fitting the parameter distribution and the all-particle energy spectrum~\cite{icetop-spectrum}.
No conclusion on the proton spectrum ``knee'' was drawn. GRAPES collaboration used a similar method to measure the proton spectrum below the knee from 0.05 to 1.3 PeV~\cite{GRAPES-3:2024mhy}.

LHAASO is a dedicated instrument designed for the purpose of primary particle identification by using a hybrid technique for detecting air showers~\cite{LHAASO-Design,LHAASO-detectors}.
The air shower parameters recorded by LHAASO include the lateral distributions of secondary particles, muon contents and the Cherenkov image of the air shower in the atmosphere.
Those high-precision measurements are achieved by three sub-arrays. The square kilometer array~(KM2A) is specially designed to have a large active muon detector with the record filling factor of 4\%~\cite{LHAASO-science-book2,LHAASO-science-book1}. It allows a precise measurement of muon content with high statistics and full containment for showers at energies as high as 100 PeV.
The large physical size of the array collects sufficient statistics which permits very strict quality cuts in event selection. The flux measurement at every energy in the spectrum has the statistical error less than 10\%. The wide field-of-view (FoV) Cherenkov telescope array (WFCTA)~\cite{LHAASO-science-book1} is designed to have both a large FoV of 16$^\circ\times$16$^\circ$ and high-resolution pixels of 0.5$^\circ$  that enables fine imaging of the shower in the development in the atmosphere. The depth of the shower maximum is measured with the resolution of 40 g/cm$^2$~\cite{LHAASO-XmaxRec}. Importantly, the imaging technique provides a high resolution shower energy measurement with a Gaussian resolution of 15\% throughout the range from 0.1 to 10 PeV covering the knee. 
They are located  4410~m above sea level, an altitude close to the atmospheric depth of the shower maximum for vertical events at $\sim$1 PeV, where the shower fluctuations reach their minimum. the tilting elevation angle of the telescopes allows selection of suitable atmospheric depths to optimize the shower measurements in different energy ranges.
All of these features make LHAASO an ideal instrument for performing a precise measurement of the proton spectrum.

\section{Hybrid measurements of air showers}
\begin{figure}[htpb]
  \centering\includegraphics[width=0.95\linewidth]{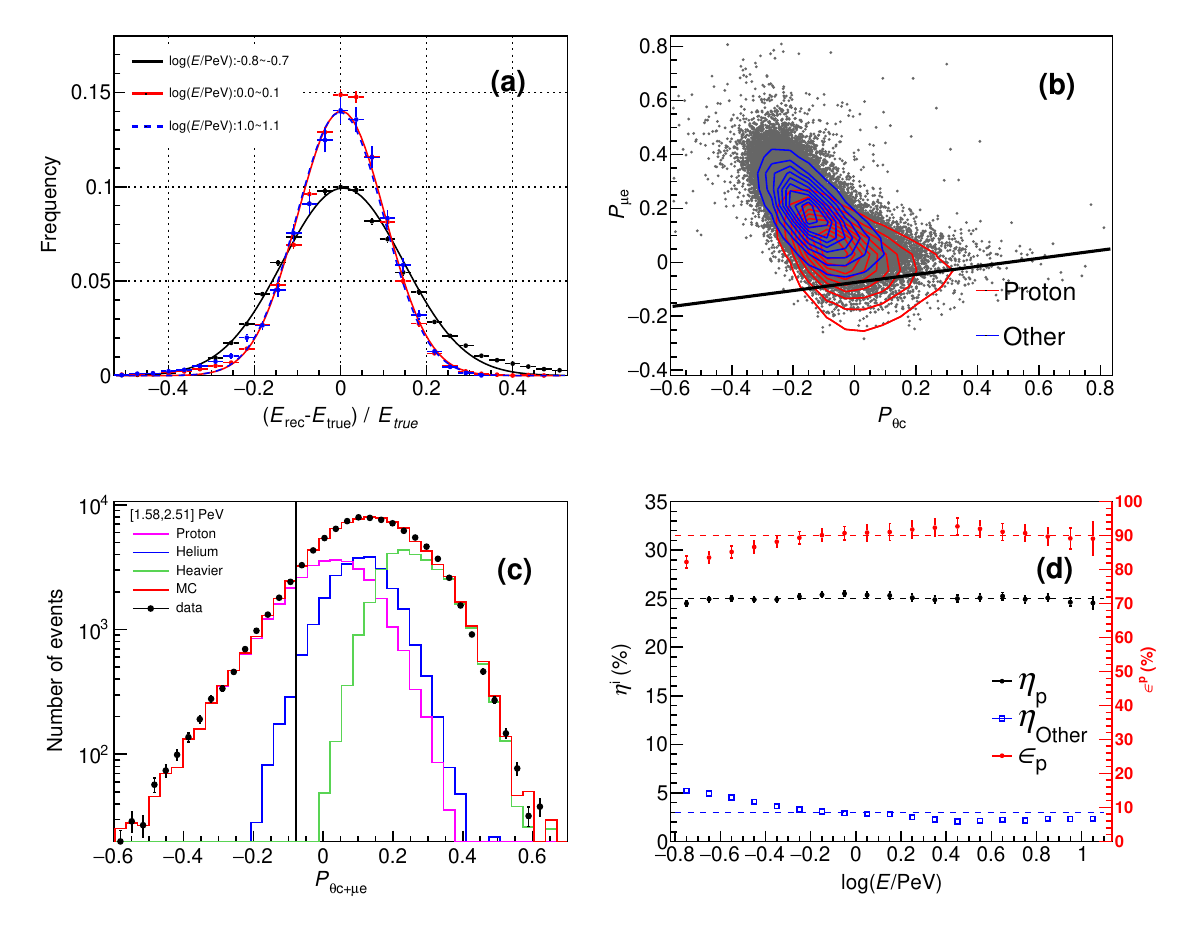}
    \caption{{Energy reconstruction, proton selection principle and performance.
  (a) The energy resolution functions for events with reconstructed energy from three energy bins. They are symmetric and well fitted with Gaussian functions, with the systematic bias less than 1\% and $\sigma$-parameter $\sim$14\%~[$\log(E/\mathrm{PeV})$=$-0.8$ to $-0.7$], $\sim$10\%~[$\log(E/\mathrm{PeV})$=0.0 to 0.1] and $\sim$10\%~[$\log(E/\mathrm{PeV})$=1.0 to 1.1], respectively.
 (b) Simulated event distributions in the two-dimensional parameter space ($P_{\mathrm{\theta c}}$, $P_{\mathrm{\mu e}}$).
 \textcolor{black}{$P_{\mathrm{\theta c}}$ is a component sensitive variable related to shower maximum measured by WFCTA, while $P_{\mathrm{\mu e}}$ is a component sensitive variable related to muon content measured by KM2A. For detailed descriptions, please refer to the following text and \textcolor{black}{Supplementary material}.}
 Many proton events (red contours) are clearly separated from the other events (blue contours).
 The gray points show the scatter plot of the association between $P_{\mathrm{\theta c}}$ and $P_{\mathrm{\mu e}}$ for heavier components.
 The black solid line indicates the selection criterion for protons.
 The events to the lower right of the black line are those retained after selection.
 (c) Distributions of $P_{\mathrm{\theta c+\mu e}}$ for events with energy between \textcolor{black}{1.58 and 2.51 PeV}. The experimental data (black dots) are shown together with simulated protons (pink histogram), helium (blue histogram), heavier components(green histogram), and \textcolor{black}{their sum which is marked as MC (red histogram). The sum of the simulated events fits the data well}. The simulation events are based on the EPOS-LHC model. The black line represents the proton selection criterion. It can be seen that, after selection, events heavier than helium (CNO+MgAlSi+Iron) are almost completely excluded, allowing the contamination of proton events to be neglected.
 (d) The corresponding selection efficiency of protons~(black dots) and other components (blue squares), as well as the purity of protons (red dots), as a function of energy. The three dashed lines at 25\%~(\textcolor{black}{black}), 90\%(red), and 3\%(blue), serve as their reference values, respectively.}
  }
  \label{fig:protonSelectResult}
\end{figure}

KM2A and WFCTA are used in the hybrid measurements of showers.
The electromagnetic particle detectors (EDs) in the KM2A array are scintillator counters that  measure the number of secondary particles passing through them and the arrival time of those particles at various distances from the shower core. The shower front is constructed, allowing for the retrieval of shower core position and arrival direction with resolutions of 6.5~m and 0.4$^\circ$ \textcolor{black}{at 0.158 PeV, improving to 1.5~m and 0.1$^\circ$ at 3 PeV}, respectively.
The other array in KM2A, muon detectors (MDs) buried 2.5 m beneath the surface count the number of muons passing through them. On average, even in the lowest energy showers used in the analysis, over 1000 muons \textcolor{black}{were} recorded, which offers highly precise measurements of the muon content. 
Images of the air showers are recorded by imaging atmospheric Cherenkov telescopes in WFCTA simultaneously. The total number of Cherenkov photons in the image is a good shower energy estimator for well-developed showers which pass their maxima before reaching the telescopes. In order to select those showers, all telescopes are tilted with the main optical axis to  45$^\circ$ in elevation thus $\sim$850 g/cm$^2$ depth of the atmosphere used in the shower development.
Combining the 
muon content and the number of photons together, reconstruction of the shower energy is further optimized  by minimizing the systematic bias below 1\%  throughout the energy range and a resolution of 10\% above 1 PeV. 
Fig.~\ref{fig:protonSelectResult}a shows the energy resolution functions for events in three energy intervals. As the energy of the shower is known, the muon content and depth of shower maximum can be normalized to be independent of energy. As a result, the correlations with the mass of primary particle become explicitly manifest. This enables identification of the primary particles. More details are provided in \textcolor{black}{Supplementary material}.

The data were collected simultaneously by KM2A and WFCTA from October 2021 to April 2022 on clear nights only. The total observation time for each telescope during the hybrid observations with KM2A was approximately 900 h.
Approximately $9.4\times 10^{6}$ events with energy between 0.158 and 12.6 PeV survived the quality cuts. The selection efficiency is 100\% within the effective aperture of $A_{\mathrm{eff}}\sim$ \textcolor{black}{75000 m$^2$ sr} for the hybrid observation.
For details of the analysis, see the corresponding descriptions in \textcolor{black}{Supplementary material}.

\section{Proton sample identification}
Showers induced by heavier nuclei have larger muon contents, which are roughly proportional to $A^{0.1}$, where $A$ is the atomic number of the nucleus~\cite{EASmodel-Matthews}.
Among ground-based arrays of comparable scale, LHAASO has the largest area of muon detectors, with a total area of \textcolor{black}{40000} m$^2$, which is evenly distributed throughout the entire array. It can effectively and evenly measure the muon contents in the EAS whose cores well contained in KM2A. Given the energy of a shower, the number of secondary particles is sensitive to the primary mass.  
The ratio $N_\mu/N_\mathrm{e}^{0.82}$ is found not only to have enhanced sensitivity to the primary mass, but also to be independent of the shower energy.
Here, $N_\mu$ and $N_\mathrm{e}$ are the number of muons and electromagnetic particles within the ring between 40 and 200 m from the shower axis.
$P_{\mathrm{\mu e}}\propto\log_{10}\frac{N_\mu}{N_\mathrm{e}^{0.82}}$ is constructed as one of the composition-sensitive parameters.
For a given  energy, proton-induced showers penetrate deeper in the atmosphere in the  longitudinal development compared with  those induced by heavy nuclei~\cite{EASmodel-Matthews}, thus proton-induced showers reach the maximum at deeper depth $X_{\mathrm{max}}$.  
The greater the maximum depth of the EAS, the larger the angle at which the photons emitted from that depth enter the telescope.
This is reflected in the Cherenkov image as an increase in the angular distance between the centroid of the image \textcolor{black}{and} the shower direction~($\theta_\mathrm{c}$).
However, $\theta_\mathrm{c}$ also increases with the shower energy and \textcolor{black}{the perpendicular distance between the telescope and the shower axis} ($R_\mathrm{p}$).
We construct $P_{\mathrm{\theta c}}$, based on the measurements of $\theta_\mathrm{c}$ and $R_\mathrm{p}$ with the dependence of energy removed, as a composition-sensitive parameter.
The separation between protons and other species in the distributions of those parameters clearly shows the capability of identifying protons out of all showers simulated with a mixture of all species. Detailed discussion is available in \textcolor{black}{Supplementary material}. In Fig.~\ref{fig:protonSelectResult}b, we show event distributions in the 
two-dimensional parameter space, and the group of proton-induced showers is separate from others.
It is found that a simple linear combination of the two parameters, namely $P_{\mathrm{\theta c+\mu e}}$, has optimized separation in the distributions as shown in Fig.~\ref{fig:protonSelectResult}c in the given energy range. The simulation matches measured data very well. An optimized energy-dependent cut that keeps events with small values of the parameter $P_{\mathrm{\theta c+\mu e}}$ is applied to all events to obtain a sample of selected events that are mostly protons with only a small contamination of helium and almost no heavier components. Fig.~\ref{fig:protonSelectResult}d indicates that the average 89.3\% purity for proton samples is achieved with the proton survival probability of 25\%, which remains constant regardless of energy according to the simulated events.
More details are shown in \textcolor{black}{Supplementary material}.

\section{The proton energy spectrum}
The CR proton energy spectrum $F(E)$ is calculated as follows
\begin{equation}
\begin{aligned}
    F(E) &=\frac{\Delta N_{\mathrm{sel}}(E)\cdot \epsilon_{\mathrm{P}}}{\Delta E\cdot A_{\mathrm{eff}}\cdot T\cdot \eta_\mathrm{P}},
    \label{equation:spectrum}
\end{aligned}
\end{equation}
where $\Delta N_{\mathrm{sel}}(E)$ is the number of selected events in the energy bin $\Delta E$ in the total exposure time $T$,
$A_{\mathrm{eff}}$ is the effective aperture, and
$\eta_\mathrm{P}$ and $\epsilon_\mathrm{P}$ are the selection efficiency of the proton events out of assumed input proton spectrum  and purity of proton events among the total selected events, respectively.
The  purity is a function of the proton energy.
\textcolor{black}{In the simulations, we employed the developed procedure to recover the proton energy spectra for four composition models to test the robustness of the analysis method.
The four models predict significantly different mixtures of proton, helium, and other components as a function of energy. For instance, the differences in the flux ratio of protons to helium can be as large as 50\% between models.
Using our developed analysis procedure, the reconstruction of the proton energy spectra for these four composition models can achieve an accuracy within 7\% below 3 PeV. See Supplemental material for further discussions.}

\begin{figure*}[htpb]
  \centering\includegraphics[width=0.55\linewidth]{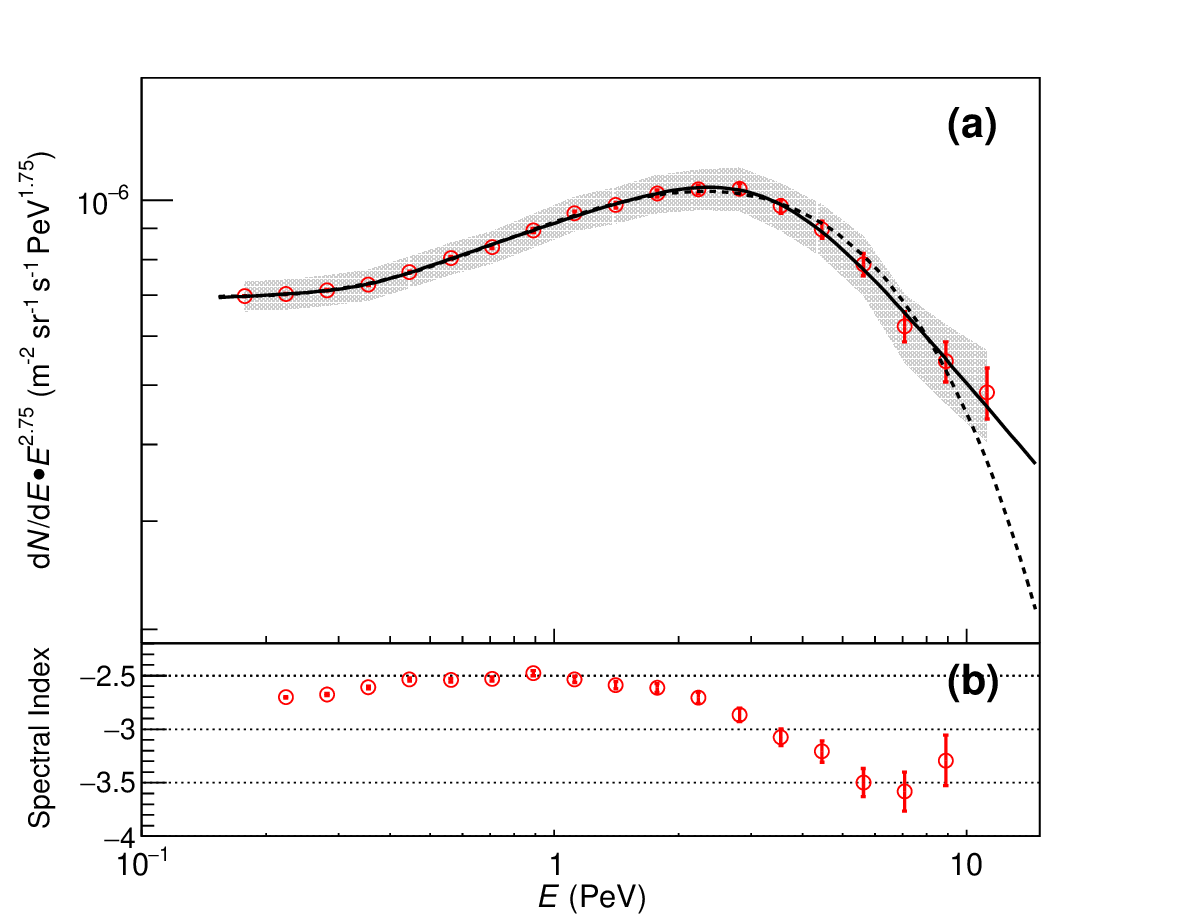}
    \caption{CR proton energy spectrum measured by LHAASO.
 (a) The proton flux multiplied by $E^{2.75}$ as a function of energy. The error bars indicate the statistical uncertainties and the shaded band indicates the systematic uncertainties. EPOS-LHC is the hadronic interaction model used in the figure.
 The solid and dashed lines represent the best fitting results  using Eq.~(\ref{formula:protonSpectrum-FitSBPL-maintext}) for three power-law components and for two power-law components with an exponential cut-off feature, respectively.
 (b) The local spectral index as a function of energy. The spectral indices are obtained using adjacent three data points fitted with a power-law function form. This indicates a slight hardening with $\Delta\gamma\sim0.2$ and a gradual softening structure (``knee'') with $\Delta\gamma\sim-1$. } 
  \label{fig:protonSpectrumIndex}
\end{figure*}

The proton energy spectrum measured by the LHAASO is calculated according to Eq.~(\ref{equation:spectrum}) and shown in Fig.~\ref{fig:protonSpectrumIndex}a.
All the relevant values of the proton spectrum including $\Delta N_{\mathrm{sel}}(E)$ are listed in the Table~S2 (online) in \textcolor{black}{Supplementary material}.

The systematic uncertainty is represented by the shaded area in Fig.~\ref{fig:protonSpectrumIndex}a.  It is itemized as follows:
\textcolor{black}{(1)} The calibration uncertainty of a single particle detected by ED and MD is less than 2\%~\cite{KM2A-ED-calibration} and 0.5\%~\cite{KM2A-MD-calibration}, respectively.
\textcolor{black}{(2)} The absolute number of photons measured by the pixel of the camera of WFCTA telescopes is calibrated using LEDs with an uncertainty of 2.6\%~\cite{WFCTA-LED-calibration}. 
\textcolor{black}{(3)} The uncertainty due to the corrections for variations in the total overburden grammage, absolute humidity in the air and light background level is estimated by dividing the data into two batches according to those three conditions.
\textcolor{black}{For overburden grammage, we divided the data into two groups based on whether the atmospheric pressure is greater than 595 hPa or less than this value. For absolute humidity, we divided the data into two groups based on whether the absolute humidity is greater than $1.53\times10^{-6}\mathrm{g/cm^{3}}$ or less than this value. For background light pollution, we divided the data into two groups based on whether there is moonlight or no moonlight.}
The largest differences between the spectra obtained from the two batches of data, \textcolor{black}{grouped according to the three conditions}, are 1\%, 1\% and 2\%, respectively.
\textcolor{black}{(4)} The systematic uncertainty due to assumptions of the fluxes of heavier nuclei, which is mainly due to the ratio of proton and helium fluxes $H/He$, is estimated by recovering the input proton spectrum for four ``composition models''. The discrepancy between the reconstructed energy spectrum and the model provided spectrum is found to be less than 7\% below 3 PeV.
It is worth pointing that this uncertainty will be diminished significantly once the $H/He$  is determined by future LHAASO measurements. Details are available in \textcolor{black}{Supplementary material}.

The energy spectra obtained based on three hadronic interaction models, QGSJETII-04~\cite{QGSJETII-04}, EPOS-LHC~\cite{EPOS-LHC}, and SIBYLL 2.3d~\cite{SIBYLL2.3d} in the shower simulation are presented in Fig.~\ref{fig:protonSpectrum}a. The difference between the three spectra \textcolor{black}{does} not exceed 17\%.
Since the uncertainty introduced by the hadronic models dominates among all systematic uncertainties,
comparison between the distributions of sensitive parameters, such as $P_{\mathrm{\theta c+\mu e}}$, between the measured and simulated events is useful to test the reliability of hadronic models.
The values of the proton energy spectrum, along with the corresponding statistical and combined systematic uncertainties,
are listed in Table~S2 (online) for different hadronic models in \textcolor{black}{Supplementary materials}.

\subsection {Spectrum fitting and discussion on hardening and proton knee}

To examine the shape of the spectrum as it varies with energy, the local spectral index determined using three adjacent points is also plotted in Fig.~\ref{fig:protonSpectrumIndex}b as a function of energy.
It is seen that in addition to the well-known ``knee'' structure, we have also observed the \textcolor{black}{hint of} a hardening of the energy spectrum below the ``knee''.

The best fit of the spectrum over the entire energy range from 0.158 to 12.6 PeV comes with three power laws, i.e., a combination of two power-law breaks. The functional form reads as
\begin{scriptsize}
\begin{equation}
    F(E) = F_0 \left(\frac{E}{\textcolor{black}{0.2}~\mathrm{PeV}}\right)^{\gamma_1}\left(1+\left(\frac{E}{E_{\mathrm{h}}}\right)^{1/w_1}\right)^{(\gamma_2-\gamma_1)w_1}\left(1+\left(\frac{E}{E_{\mathrm{k}}}\right)^{1/w_2}\right)^{(\gamma_3-\gamma_2)w_2}.
     \label{formula:protonSpectrum-FitSBPL-maintext}
\end{equation}
\end{scriptsize}

The fitting results are presented in Table~\ref{table:protonSpectrum-FitSBPL-maintext}.
The second type of error is the systematic error, which is derived by comparing the fitting results of the energy spectra under various conditions, such as different air pressures, selection efficiencies, hadronic interaction models, composition models, etc. When focusing on the changes in the spectral index, all fitting results show that the change in the power-law index, $\Delta\gamma = \gamma_{2}-\gamma_{1}$, exceeds 0.14, indicating the presence of a hardening tendency in all cases. Taking into account the systematic uncertainty of fluxes at the lowest energies, the first power law component and hardening energy $E_\mathrm{h}$, which are eventually decided by the left-most 3 points, have larger uncertainty, compared to the second and third power law components.
The measurement of the knee by LHAASO is a broad feature
in the energy interval 1--10~PeV 
with a large step in spectral index $\Delta \gamma  = \gamma_{3}- \gamma_{2} \approx -1.0$.

To test the hypothesis of the softening being an exponential cut-off instead of a break in the power law spectrum, the proton spectrum is fitted with a combination of the smoothly broken power law functional form and the exponential cutoff $e^{-\frac{E}{E_{\mathrm{cut}}}}$.
As shown in Fig.~\ref{fig:protonSpectrumIndex}a, neither the statistics nor the energy coverage of the measurement allows a conclusive choice between scenarios.

Below the ``knee'', the  proton energy spectrum is much harder than the all-particle energy spectrum~\cite{LHAASO-allparticle}, indicating that the proportion of proton events is rapidly increasing before bending down in the knee. The knee energy of the proton spectrum ($E_{\mathrm{k}}$$=$3.3 PeV) is essentially the same as that of the all-particle energy spectrum, 3.67 PeV~\cite{LHAASO-allparticle}, within statistical errors ($\sim$0.5~PeV).

\begin{table*}[ht]
\renewcommand{\arraystretch}{1.25}
\begin{center}
\begin{threeparttable}
\caption{Fitting results of proton spectrum using Eq.~(\ref{formula:protonSpectrum-FitSBPL-maintext}). }
\setlength{\tabcolsep}{3pt}
\begin{tabular}{ccccccccc}
\hline
Model & $F_{0}$   & $E_{\mathrm{h}}$  & $\gamma_{1}$ & $\gamma_2$ & $w_1$ & $E_{\mathrm{k}}$  & $\gamma_3$ & $w_2$ \\
 &  $(\mathrm{m^{-2}}$ $\mathrm{sr^{-1}}$ $\mathrm{s^{-1}}$ $\mathrm{PeV^{-1})}$ & (PeV) &  &  &  &  (PeV) &  &  \\
\hline
EPOS-LHC & (5.85 & 0.34 & $-2.71$ & $-2.51$ &  0.12 & 3.3 & $-3.5$ & 0.27 \\
 & $\pm$0.02$\pm$0.17)$\times 10^{-5}$ & $\pm$0.02$\pm$0.04 & $\pm$0.02$\pm$0.08 & $\pm$0.03$\pm$0.06 &  $\pm$0.08$\pm$0.16 & $\pm$0.4$\pm$0.5 & $\pm$0.2$\pm$0.2 & $\pm$0.07$\pm$0.09 \\
QGSJETII-04 & (6.45 & 0.37 & $-2.79$ & $-2.54$ &  0.33 & 3.4 & $-3.6$ & 0.31 \\
 & $\pm$0.23$\pm$0.36)$\times 10^{-5}$ & $\pm$0.06$\pm$0.10 & $\pm$0.07$\pm$0.12 & $\pm$0.14$\pm$0.09 &  $\pm$0.39$\pm$0.29 & $\pm$0.5$\pm$0.7 & $\pm$0.3$\pm$0.3 & $\pm$0.13$\pm$0.12 \\
 SIBYLL & (5.52 & 0.35 & $-2.76$ & $-2.52$ &  0.14 & 3.3 & $-3.6$ & 0.31 \\
 & $\pm$0.02$\pm$0.18)$\times 10^{-5}$ & $\pm$0.02$\pm$0.02 & $\pm$0.02$\pm$0.07 & $\pm$0.04$\pm$0.05 &  $\pm$0.08$\pm$0.15 & $\pm$0.5$\pm$0.5 & $\pm$0.2$\pm$0.2 & $\pm$0.09$\pm$0.06 \\
\hline
\end{tabular}
\label{table:protonSpectrum-FitSBPL-maintext}
\begin{tablenotes}
\item $F_0$ is the normalization constant, representing the proton flux at \textcolor{black}{0.2} PeV. 
\textcolor{black}{$E_\mathrm{h}$ and $E_\mathrm{k}$ are the energy values at which the proton energy spectrum exhibits a hardening and a knee, respectively.} $\gamma_1$, $\gamma_2$, and $\gamma_3$ are the spectral indices of the energy spectrum below hardening, between hardening and the knee, and above the knee, respectively. The parameters $w_1$ and $w_2$ give the width in $\log E$ of the step in spectral index.
\end{tablenotes}
\end{threeparttable}
\end{center}
\end{table*}
The results of the spectra obtained using three hadronic interaction models are plotted in Fig.~\ref{fig:protonSpectrum}a.
The spectral features can be recognized in spite of the systematic errors due to the uncertainties in the interaction models.

\subsection{Comparison with other measurements}

\begin{figure*}[htpb]
  \centering\includegraphics[width=0.7\linewidth]{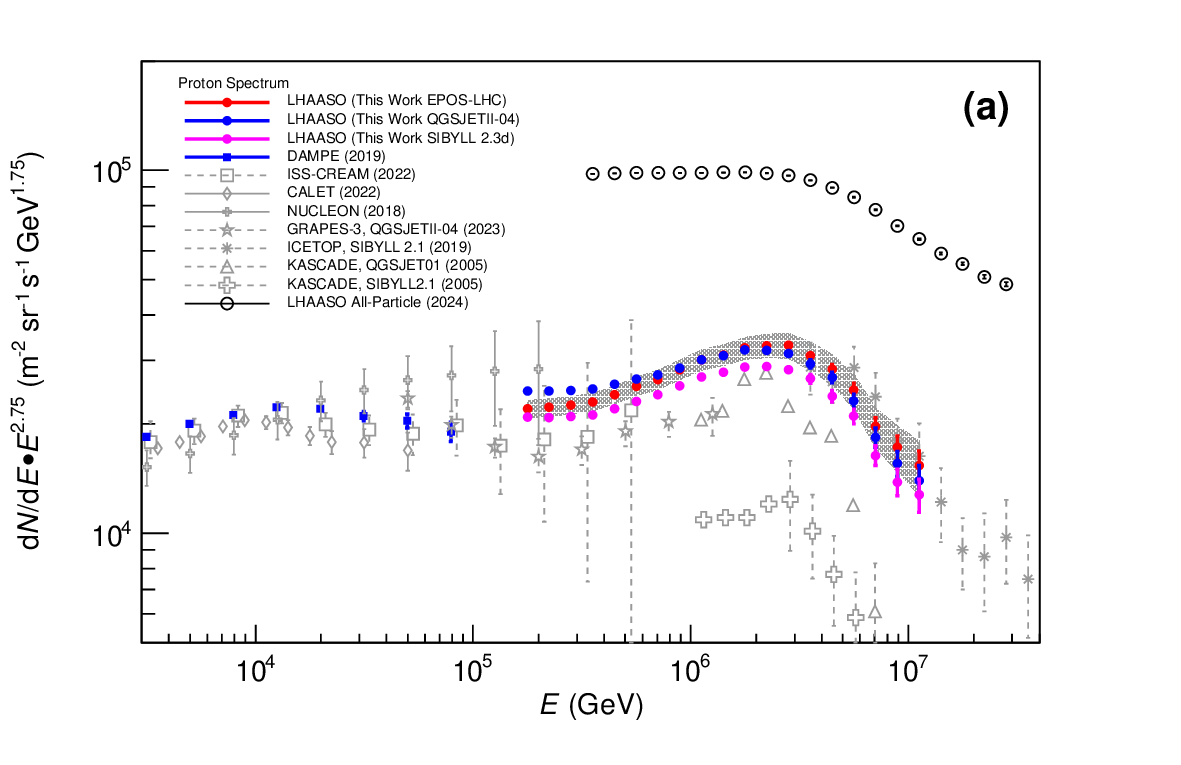}
  \centering\includegraphics[width=0.7\linewidth]{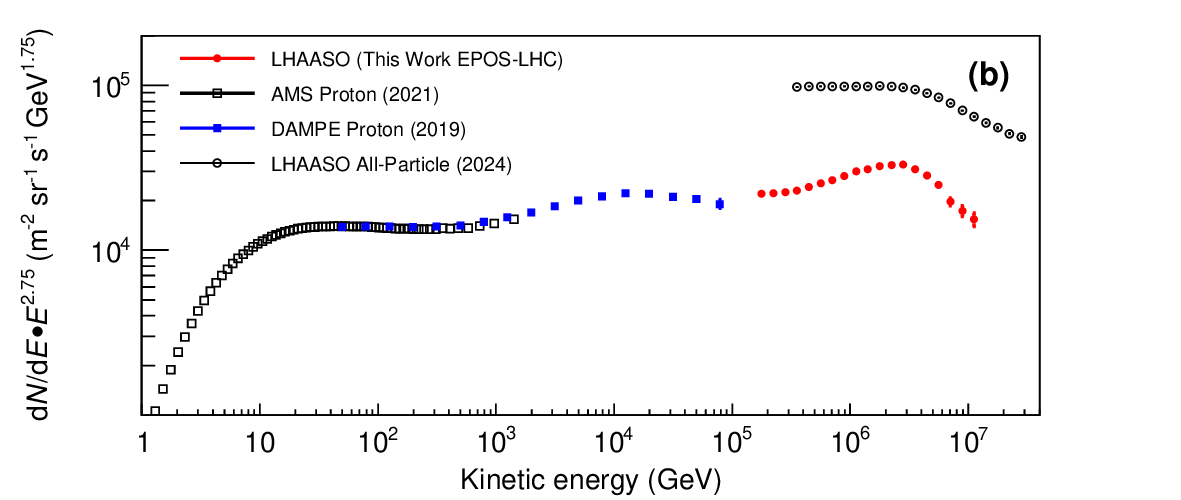}
    \caption{{ (a) CR proton spectrum from a few TeV to tens of PeV. The fluxes are multiplied by a factor of $E^{2.75}$, allowing a detailed comparison between measurements in entirely different energy domains. Proton spectra reported by the space-borne ISS-CREAM, DAMPE, CALET, and NUCLEON and ground-based GRAPES-3, ICETOP, and KASCADE detectors and the LHAASO all-particle spectrum are plotted together with the proton spectrum measured in this work by LHAASO. Here we show three proton energy spectra based on different hadronic models. All error bars represent the statistical errors. The shaded band represents the overall systematic uncertainty, excluding that introduced by the hadronic interaction models. The center of the shaded band corresponds to the results from the EPOS-LHC model. The systematic uncertainty introduced by the hadronic interaction models can be referenced by the differences among the three energy spectra.
 (b) To clearly see how those measurements are connected over a wider energy range starting from 1 GeV, the spectra by AMS-02, DAMPE and LHAASO (proton spectrum and all-particle spectrum) \textcolor{black}{are plotted}. AMS-02 and DAMPE have a good overlap around $10^{3}$ GeV, and the gap between DAMPE and LHAASO around 0.1 PeV should be covered by the two experiments shortly, allowing a complete description of the spectrum without suffering from a relative energy scale difference between experiments.  
  }
  }
  \label{fig:protonSpectrum}
\end{figure*}

The proton spectrum as measured by LHAASO  is compared with other  ground-based detectors measurements by GRAPES-3~\cite{GRAPES-3:2024mhy}, ICETOP~\cite{icetop-spectrum}, KASCADE~\cite{kaskade-spectrum-00,kaskade-spectrum-01,kaskade-spectrum-02} and LHAASO all-particle spectrum~\cite{LHAASO-allparticle}, as well as the space-borne experiments  AMS~\cite{AMS2021-spectrum}, DAMPE~\cite{DAMPE:2019gys}, ISS-CREAM~\cite{Choi:2022aht}, CALET~\cite{CALET:2022vro} and NUCLEON~\cite{Atkin:2018wsp}, see Fig.~\ref{fig:protonSpectrum}a.

For the proton spectrum measured by LHAASO, the spectral index is approximately $-2.51$ within the energy range of 0.3-1.5 PeV for three interaction models. Taking into account all possible uncertainties, which are 
approximately 0.08 in the spectral index, the spectrum is significantly harder than the one exhibiting the so-called ``softening'' feature, 
at energies above the break
around $13.6^{+4.1}_{-1.8}$~TeV(=1/1000 PeV) according to DAMPE~\cite{DAMPE:2019gys}, $9.3^{+1.4}_{-1.1}$~TeV according to CALET~\cite{CALET:2022vro}, and ($9.0\pm1.3$)~TeV according to ISS-CREAM~\cite{Choi:2022aht}.
The spectral indices are measured as $-2.85 \pm 0.07$ by DAMPE~\cite{DAMPE:2019gys}, $-2.89 \pm 0.07$ by CALET~\cite{CALET:2022vro}, and $-2.82 \pm 0.02$ by ISS-CREAM~\cite{Choi:2022aht}.
These measurements imply that a ``hardening'' must occur at some energy between 0.1  and 0.4 PeV.
Identifying the exact $E_\mathrm{h}$ is an important task requiring the sustained efforts from both space-borne and ground-based experiments.

In Fig.~\ref{fig:protonSpectrum}b, the proton spectrum is shown over 7 decades of energy. Only the data of AMS-02, DAMPE and LHAASO with small statistical uncertainties are plotted. 
Two features have been revealed by direct measurements in space: a hardening at
0.5 TeV and a softening above 10 TeV. At higher energies ($E$$\sim$3.5 PeV), the marked softening of the ``knee'' has been observed by LHAASO. The existence of an additional
hardening feature centered at $\sim$0.1 PeV is evident, while the energy is in the specific range between direct and indirect measurements where there \textcolor{black}{is} still a clear gap between the measurements.

\section{Discussions}
The Galactic CR spectrum encodes both the initial energy distributions imparted by 
astrophysical accelerators and the modiﬁcations introduced during CR propagation through the Milky Way. Especially, the detailed spectral features of the individual CR species, protons in particular, 
provide reliable constraints on theoretical models.
\textcolor{black}{For instance, several possible interpretations of the spectra of protons and other nuclei are presented in terms of combinations of multiple populations of CR sources, following the approach of Gaisser et al~\cite{Gaisser:2013bla}.} There are more recent attempts~\cite{Liu:2018fjy,Fang:2020cru, Li:2021szb,Qiao:2022cge,Zhang:2022pzt} to explain the hardening and softening feature below 0.1 PeV.
LHAASO's measurement reveals that the ``knee'' appears as a broad hump, peaking around 3 PeV, in the spectrum. The rising part, i.e. the hardening of the proton spectrum presented here, may indicate that the trend of softening is overtaken by a new component, probably associated with the emissions of a class of accelerators different from those that dominate the flux below 0.1 PeV~\cite{Prevotat:2025ktr, Zhang:2025tew}.

Given the theoretical challenges of 
accelerating protons to PeV energies particularly in supernova remnants, e.g.~\cite{Bell:2013kq}, it has been
proposed that the knee region is instead dominated by heavy nuclei, implying that protons are only 
accelerated effectively up to $\sim$0.1 PeV. This hypothesis has been widely discussed in the Ref.~\cite{Sveshnikova:2003sa}. LHAASO now unambiguously rules out this scenario by collecting large samples of proton events at energies higher than 10 PeV. It is natural to link this result to the dozens of newly discovered PeVatrons as the ultra-high energy gamma ray sources by LHAASO~\cite{LHAASO-catalogue,LHAASO-nature-12}.

The central issue is whether the spectral features observed above 0.1 PeV result from changes in CR propagation or the emergence of a new PeVatron 
component.
In the former case, the implication is that the Galactic accelerators are operating efﬁciently across the 
entire CR energy range, up to PeV energies. This scenario is energetically plausible for supernova 
remnants and stellar clusters, which can provide the required power of approximately $10^{41}$ erg/s to sustain the observed CR ﬂux~\cite{Strong:2010pr}.
On the other hand, microquasars or other super-Eddington black hole 
binaries represent alternative candidates for the second (PeVatron) component. They require a relatively modest injection power of the order of $\sim 10^{39}$ erg/s or less~\cite{Zhang:2025tew, LHAASO:2024psv, Alfaro:2024cjd, Wang:2025yqy}.

\section*{Conflict of interest}
The authors declare that they have no conflict of interest.

\section*{Acknowledgements}
We would like to thank all staff members who work at the LHAASO site at 4400 meters above sea level year round to maintain the detector and keep the water recycling system, electricity power supply and other components of the experiment operating smoothly. We are grateful to the Chengdu Management Committee of Tianfu New Area for the constant financial support for research using LHAASO data. We appreciate the computing and data service support provided by the National High Energy Physics Data Center for the data analysis in this paper.
This work was supported by the National Key R\&D Program of China (2024YFA1611401, 2024YFA1611402, 2024YFA1611403, and 2024YFA1611404), the National Natural Science Foundation of China (12393851, 12393852, 12393853, 12393854, 12205314, 12105301, 12305120, 12261160362, 12105294, U1931201, 12375107, 12275280, 12105293, 11905240, 12375106, and 12261141691), Innovation Project of IHEP (E25451U2), Sichuan Province Science Foundation for Distinguished Young Scholars (2022JDJQ0043), the Youth Innovation Promotion Association of the Chinese Academy of Sciences (CAS YIPA) (2023019). We are grateful to the Institute of Plateau Meteorology, CMA Chengdu to maintain meteorological data, and Thailand's National Science and Technology Development Agency (NSTDA) and National Research Council of Thailand (NRCT) under the High-Potential Research Team Grant Program (N42A650868).\\

\section*{Author contributions}
Zhiyong You and Shoushan Zhang were responsible for drafting the manuscript, while Zhen Cao, as the LHAASO Collaboration spokesperson and chief scientist, led finalizing of the manuscript and discussions with Paolo Lipari, Felix Aharonian, and Bing Theodore Zhang. Shoushan Zhang led a data analysis team including Zhiyong You, Liqiao Yin and Lingling Ma that conducted the analysis of the proton spectrum. Zhiyong You and Liqiao Yin also conducted a detailed analysis of the simulated data, while Suhong Chen and Liping Wang performed cross-checking. Shoushan Zhang also led the team including Yudong Wang and Mingjie Yang to calibrate the Cherenkov telescopes. Other authors participated in manuscript editing, data analysis, including event reconstruction, simulations, and event building with multiple components, detector calibration, and the operation and maintenance of scintillator counters, muon detectors, and Cherenkov telescopes. Several authors also participated in the construction and deployment of the detectors.

\section*{Appendix A. Supplementary material}
Supplementary data to this article can be found online at https://doi.org/10.1016/j.scib.2025.10.048.

\newpage
\onecolumn

\centerline{\Large \bf LHAASO~Collaboration}

\noindent 
Zhen Cao$^{1,2,3}$,
F. Aharonian$^{3,4,5,6}$,
Y.X. Bai$^{1,3}$,
Y.W. Bao$^{7}$,
D. Bastieri$^{8}$,
X.J. Bi$^{1,2,3}$,
Y.J. Bi$^{1,3}$,
W. Bian$^{7}$,
A.V. Bukevich$^{9}$,
C.M. Cai$^{10}$
% ...

\noindent
$^{1}$ Key Laboratory of Particle Astrophysics \& Experimental Physics Division \& Computing Center, Institute of High Energy Physics, Chinese Academy of Sciences, Beijing 100049, China\\
$^{2}$ University of Chinese Academy of Sciences, Beijing 100049, China\\
$^{3}$ TIANFU Cosmic Ray Research Center, Chengdu 610213,  China\\
$^{4}$ University of Science and Technology of China, Hefei 230026, China\\
$^{5}$ Yerevan State University, Yerevan 0025, Armenia\\
$^{6}$ Max-Planck-Institut for Nuclear Physics, Heidelberg 69029, Germany\\
$^{7}$ Tsung-Dao Lee Institute \& School of Physics and Astronomy, Shanghai Jiao Tong University, Shanghai 200240, China\\
$^{8}$ Center for Astrophysics, Guangzhou University, Guangzhou 510006, China\\
$^{9}$ Institute for Nuclear Research of Russian Academy of Sciences, Moscow 117312, Russia\\
$^{10}$ School of Physical Science and Technology \&  School of Information Science and Technology, Southwest Jiaotong University, Chengdu 610031, China\\
% ...

\newpage
\begin{center}
{\LARGE Supplementary Materials for\\
\vspace{5pt}
Precise measurements of the Cosmic Ray Proton energy spectrum in the ``knee''  region}
\end{center}

\vspace{100pt}

\noindent{\bf This PDF file includes}
\begin{itemize}
    \item[] Materials and Methods
    \item[] Supplementary Text
    \item[] Figs. S1 to S17
    \item[] Tables S1 to S3
\end{itemize}

\clearpage

\setcounter{equation}{0}
\renewcommand{\theequation}{S\arabic{equation}}

\setcounter{figure}{0}
\renewcommand{\thefigure}{S\arabic{figure}}

\setcounter{table}{0}
\renewcommand{\thetable}{S\arabic{table}}

\section*{Materials and Methods}
\vspace{5pt}
\setcounter{section}{0}
\section{Experiment description}

Large High Altitude Air Shower Observatory (LHAASO) is a complex of extensive air shower (EAS) detectors installed on Mt. Haizi (29$^\circ$21'27.6" N, 100$^\circ$08'19.6" E) at  4410 m above sea level, in Sichuan province, China.

\subsection{KM2A} 
\begin{figure}[htpb]
\centering\includegraphics[width=0.95\linewidth]{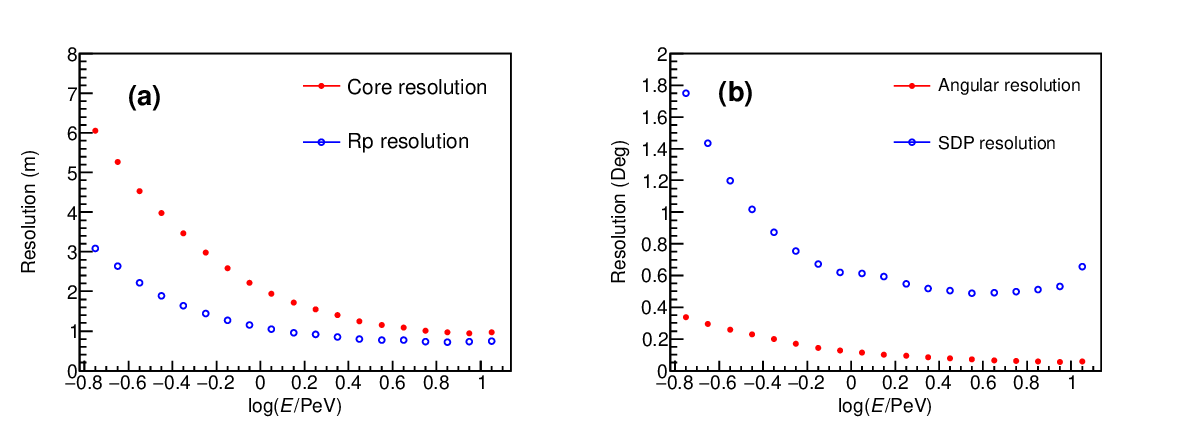}
\caption{{ 
(a) The core and direction of the shower are reconstructed with KM2A, while the shower detector plane SDP is reconstructed with WFCTA. (b) The resolutions of core, angular, telescope distance to the shower $R_\mathrm{p}$, and SDP as functions of the reconstruction energy are presented}.}
\label{fig:geo-resolution}
\end{figure}

The square kilometer array (KM2A) has a total area of 1.3 square kilometers, including 5216 electromagnetic particle detectors~(EDs) and 1188 muon detectors~(MDs).
Each ED and MD detector has a size of 1 $\mathrm{m^2}$ and 36 $\mathrm{m^2}$, respectively. The spacing between EDs is 15 m, and the spacing between MDs is 30 m.
A three-level quality control system has been established to monitor the status of detector units, the stability of reconstructed parameters and the performance of the array.
Applying this monitoring system to the data collected during the period from August 2021 to July 2023, the results show that over 98\% of the observed data are valid for physical analysis~\cite{KM2A-monitor}.
EDs measure the number and arrival time of secondary particles.
The aim is to accurately reconstruct the core position and direction of the air shower.
This is achieved by fitting the lateral distribution of secondary particles following the Nishimura–Kamata–Greisen function~\cite{KM2AcrabCPC} and the shower front.
Above 0.158 PeV, the core position resolution and directional resolution are better than 6.5 m and 0.4$^\circ$, respectively~(shown in Fig.~\ref{fig:geo-resolution}).
The MDs have a total active area of 40,000 m$^2$ with 4\% fill-factor.
The muon content in the shower is measured by MDs combining the shower geometric parameters, enabling the effective distinction of primary particle types.
It provides LHAASO with excellent $\gamma$-ray identification capability, facilitating the discovery of numerous ultra-high-energy gamma-ray sources~\cite{LHAASO-catalogue}, and also enables the measurement of the average composition $\ln A$ of CRs~\cite{LHAASO-allparticle}.
The main scientific goal of KM2A is to explore the origin of cosmic rays (CRs), accurately measure the energy spectrum of gamma-ray sources above 0.03 PeV, and provide decisive observational results on CR sources.

\subsection{Wide FoV Cherenkov Telescope Array }
\begin{figure}[htpb]
\centering\includegraphics[width=0.46\linewidth]{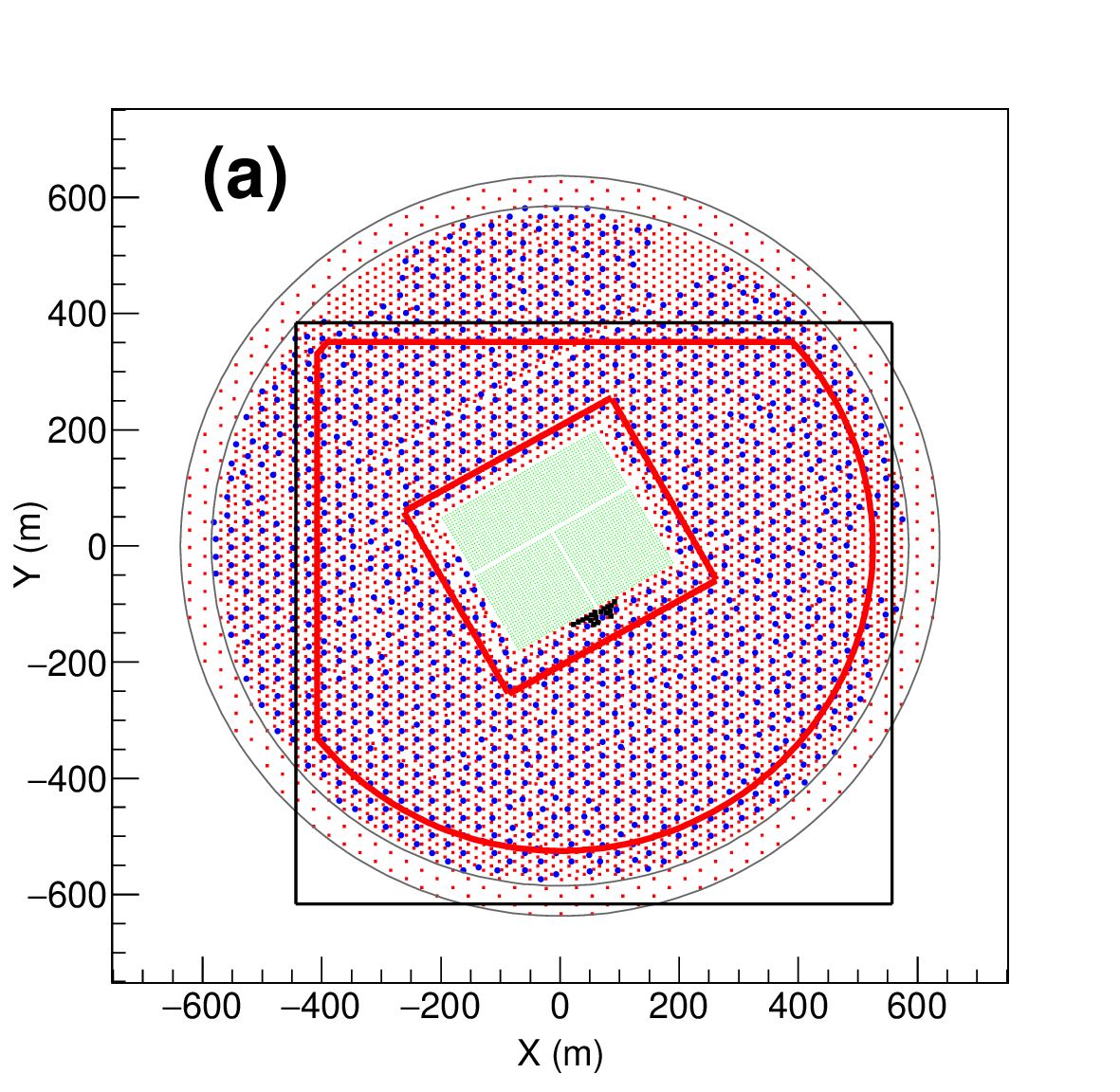}
\centering\includegraphics[width=0.46\linewidth]{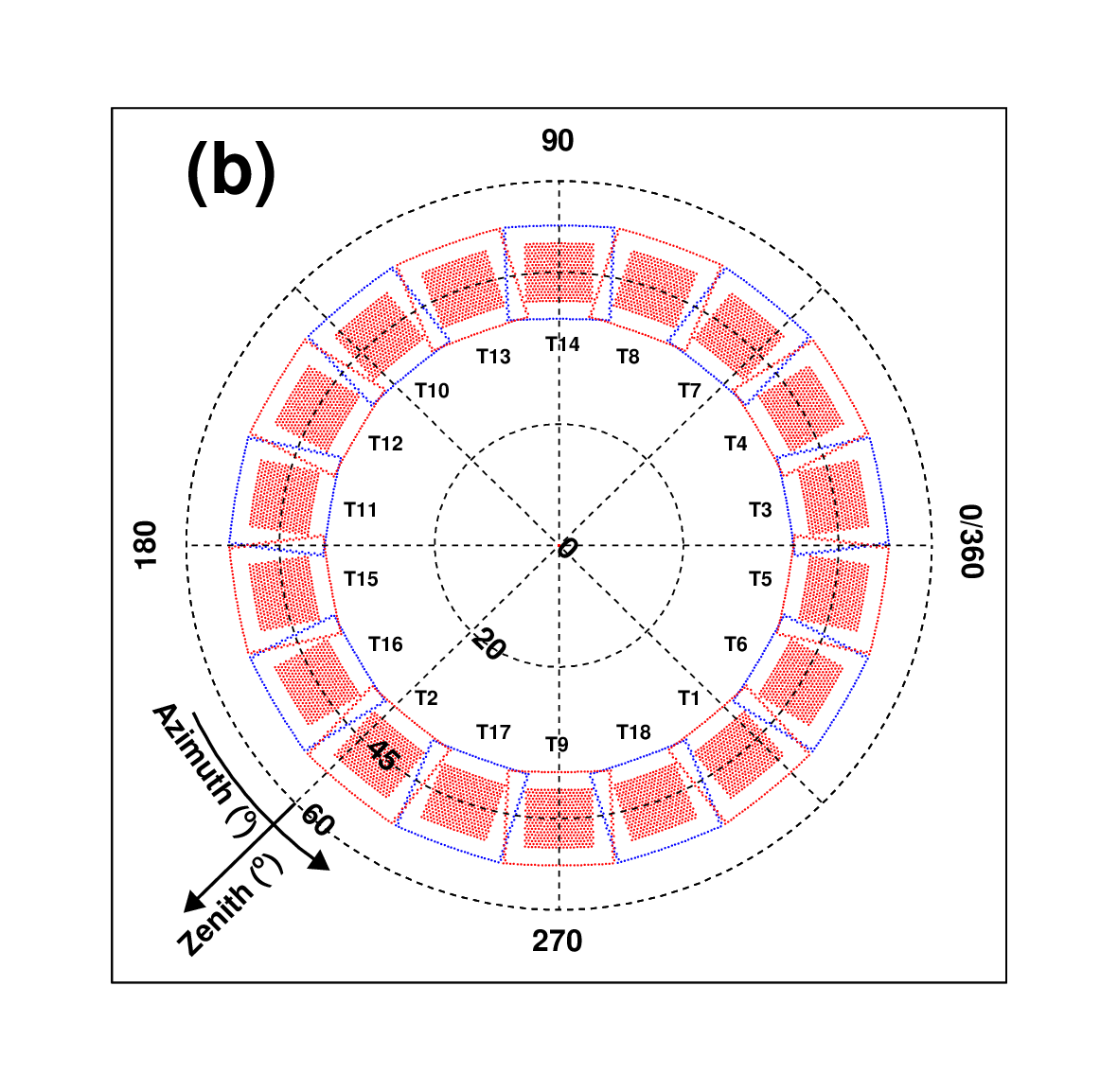}
\caption{LHAASO layout. (a): 
The WCDA is located at the center of the LHAASO detector array, surrounded by the KM2A array and 18 Cherenkov telescopes of WFCTA. 
The red solid line indicates the area for selecting the reconstructed shower core, while the black solid line represents the throwing area for the shower core in the simulation. The distance from the core selection area to the KM2A boundary exceeds 50 m, and the distance from the throwing area boundary to the core selection area is more than 30 m, ensuring accurate geometric reconstruction.
 (b): The colored dashed line represents the FoV of WFCTA. The red area represents the FoV used for selecting the centroid of the Cherenkov image in the analysis.}
\label{fig:coreselect1}
\end{figure}

The wide field-of-view (FoV) Cherenkov telescope array (WFCTA) consists of 18 wide FoV imaging atmospheric Cherenkov telescopes, which can operate in both Cherenkov light mode and fluorescence mode.
Each telescope has a 5 $\mathrm{m^2}$ spherical mirror, which reflects Cherenkov photons onto the camera plane. The distance from the camera to the mirror is 2870 mm.
The camera's photoelectric conversion devices are silicon photomultipliers (SiPMs), with a total of 1024 SiPMs arranged in a 32$\times$32 grid. Each SiPM has a field-of-view of 0.5$^\circ \times$ 0.5$^\circ$. The field of view for each telescope is $\pm8^\circ$.
The camera and mirror are both placed in a container that can adjust its elevation angle.
The elevation angle adjustment range is from 0$^\circ$ to 90$^\circ$.
The entire container is installed on a movable chassis.
Due to the mobility feature of WFCTA,  it is arranged in different array layouts in different stages to conduct hybrid observations with KM2A and water Cherenkov detector array (WCDA).
The Cherenkov image of an air shower as recorded by telescopes of WFCTA originates from Cherenkov photons emitted by the secondary charged particles in the shower. The total number of photons in the image serves as a reliable estimator of the total energy of the shower. 
Along the long axis of the elliptical image,  the number of photons increases up to its maximum and gradually decreases. This reflects the shower development feature in the \textcolor{black}{atmosphere}.  
At a given shower energy and perpendicular distance from an air Cherenkov telescope to the shower axis ($R_\mathrm{p}$), the angular offset of the brightest part of the image, or the centroid of the image, with respect to the shower arrival direction, denoted as $\theta_{\mathrm{c}}$, measures the atmospheric depth of the shower maximum, $X_{\mathrm{max}}$.
In order to allow air showers to develop well before reaching the ground, the 18 telescopes are tilted at a zenith angle of $45^\circ$ and cover the shower zenith angle range from 37$^\circ$ to 53$^\circ$, thus maintaining the high resolution of the $X_{\mathrm{max}}$ and energy. The 18 telescopes are oriented so that the FoVs cover all azimuth directions.
The data used in this work are obtained from the hybrid observations of the telescope and KM2A array.
The field of view of all telescopes is shown in Fig.~\ref{fig:coreselect1}b.

\subsection{Hybrid observation of WFCTA and KM2A}
\begin{figure}[htpb]
\centering\includegraphics[width=0.95\linewidth]{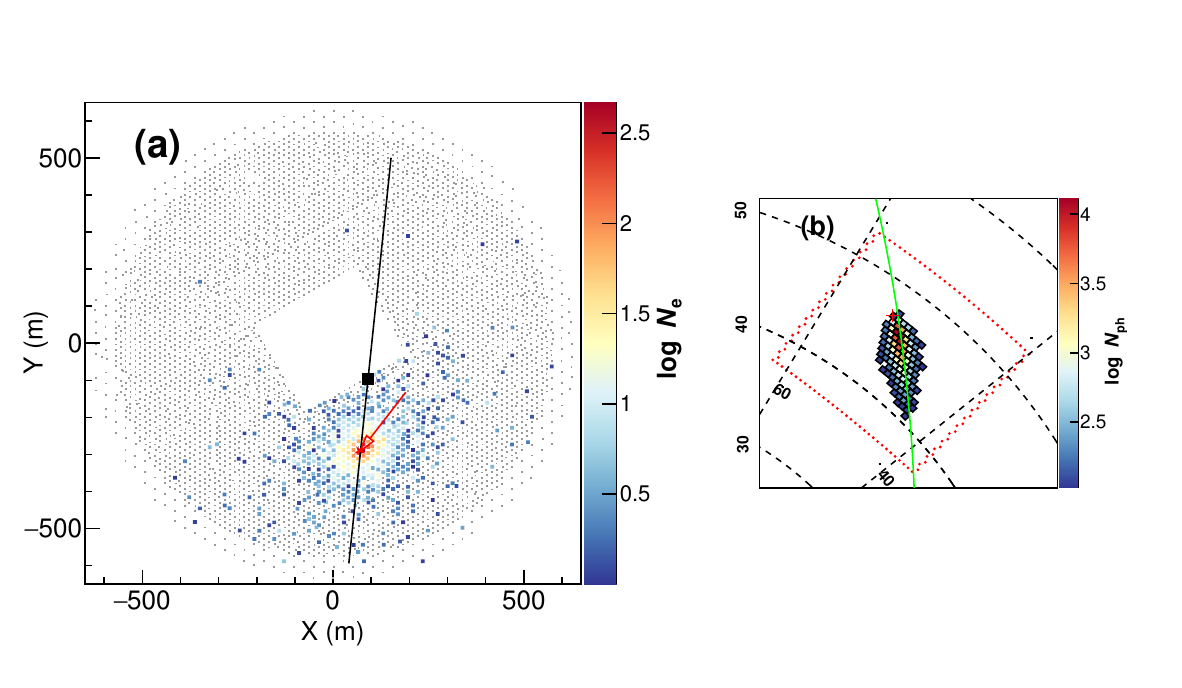}
\caption{An event measured by EDs and a Cherenkov telescope of WFCTA simultaneously. (a):  Gray squares indicate the scintillator counters of KM2A, colored according to the logarithm of the number of detected particles. The position of
the core is indicated by the red arrow, which is orientated in the arrival direction of
the primary photon. The black line is the intersection of the shower detector plane and the ground, which is consistent with the shower core.
(b): Cherenkov image recorded by a telescope of WFCTA.
The color scale shows the logarithm of the number of photoelectrons in each pixel. The main axis, indicated by the green line, of the image in the FOV of the telescope indicates the shower detector plane, which is consistent with the event direction, indicated by the red cross, reconstructed using KM2A.
Telescope FOV outlined by the red dotted lines. The dashed arcs indicate zenith angles of $30^\circ$, $40^\circ$, and $50^\circ$ from bottom to top, and dashed lines indicate azimuth angles of $40^\circ$ and $60^\circ$ counterclockwise.}
\label{fig:showEvent}
\end{figure}

WFCTA and KM2A use the same clock system, White Rabbit timing system, for independent observations.
Subsequently, joint triggering is performed offline by matching the event trigger times from both arrays. The width of the event trigger time window is 10~$\mu$s.
Fig.~\ref{fig:showEvent}a shows the distribution of ED signals in an event measured by KM2A and WFCTA simultaneously.
Fig.~\ref{fig:showEvent}\textcolor{black}{b} shows the Cherenkov image recorded in the focal plane of a telescope for the same event.
The color represents the number of photons recorded by each SiPM.
The green line, referred to as the long axis, minimizes the signal-weighted sum of squares of perpendicular angular distance of the SiPMs.
All points within the long axis, when projected into three-dimensional space, point towards the shower axis.
Two points on the long axis are used to reconstruct the normal vector of the Shower Detector Plane (SDP) composed of the telescope and the shower axis.

\subsection{Calibration}
\subsubsection{Calibration of \texorpdfstring{$N_\mu$}{} and \texorpdfstring{$N_\mathrm{e}$}{}}
\begin{figure}[htpb]
    \centering
    \includegraphics[width=0.55\linewidth]{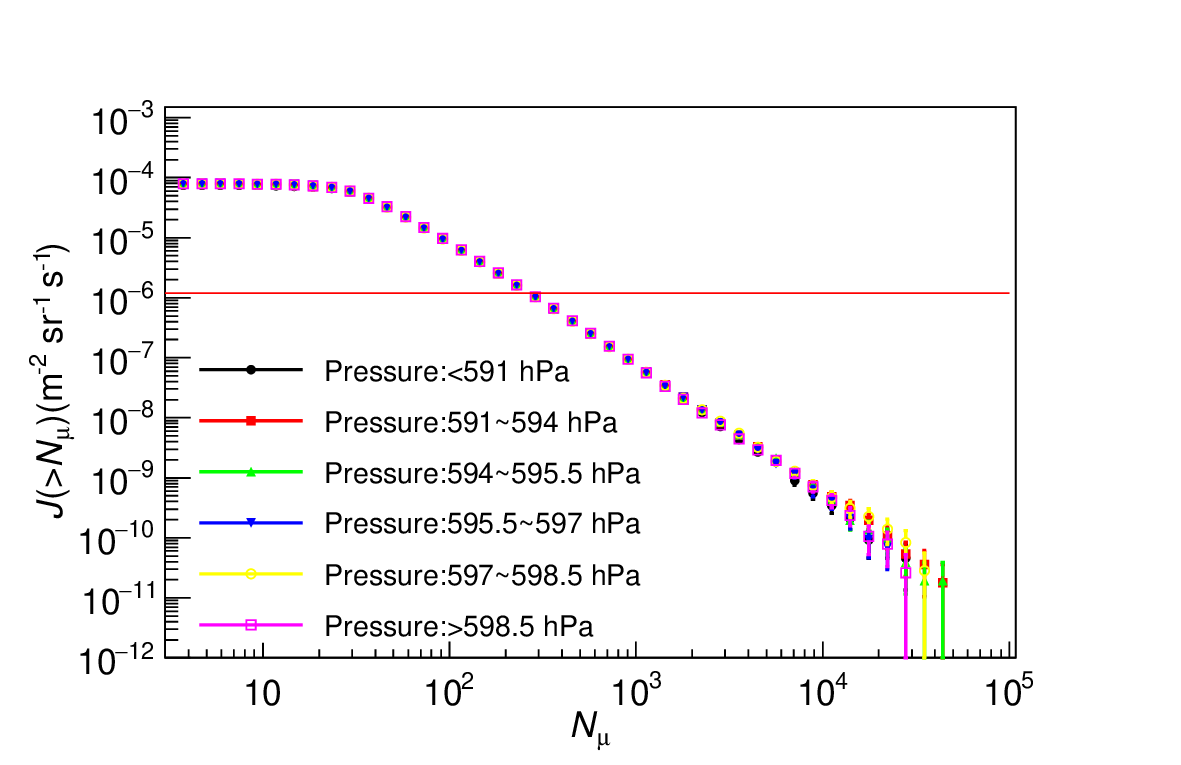}
    \caption{Muon number integral intensities for various pressure ranges, as measured by LHAASO-KM2A, are presented. The red horizontal line indicates the CIC value of $J=1.2\times 10^{-6} \mathrm{m^2\cdot sr\cdot s}$, and the error bars is statistical errors.}
    \label{fig:CICmethod}
\end{figure}

\begin{figure}[htpb]
    \centering
    \includegraphics[width=0.43\linewidth]{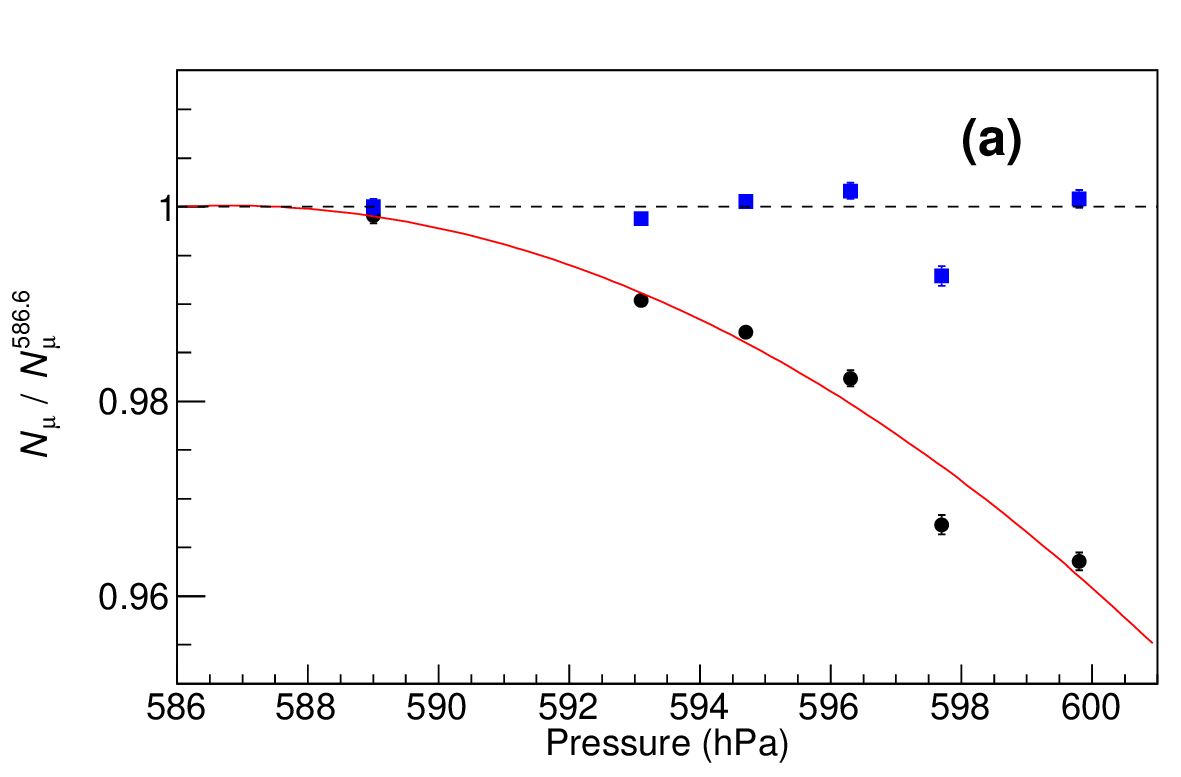}
    \includegraphics[width=0.43\linewidth]{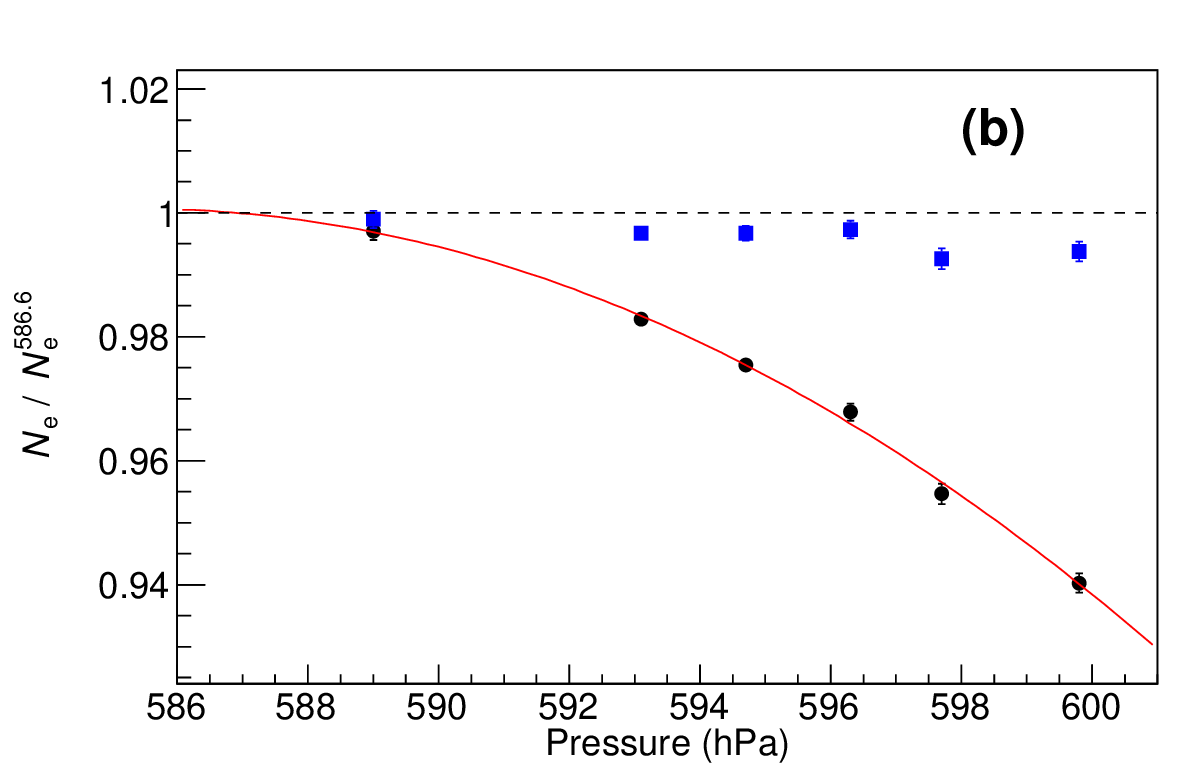}
    \includegraphics[width=0.43\linewidth]{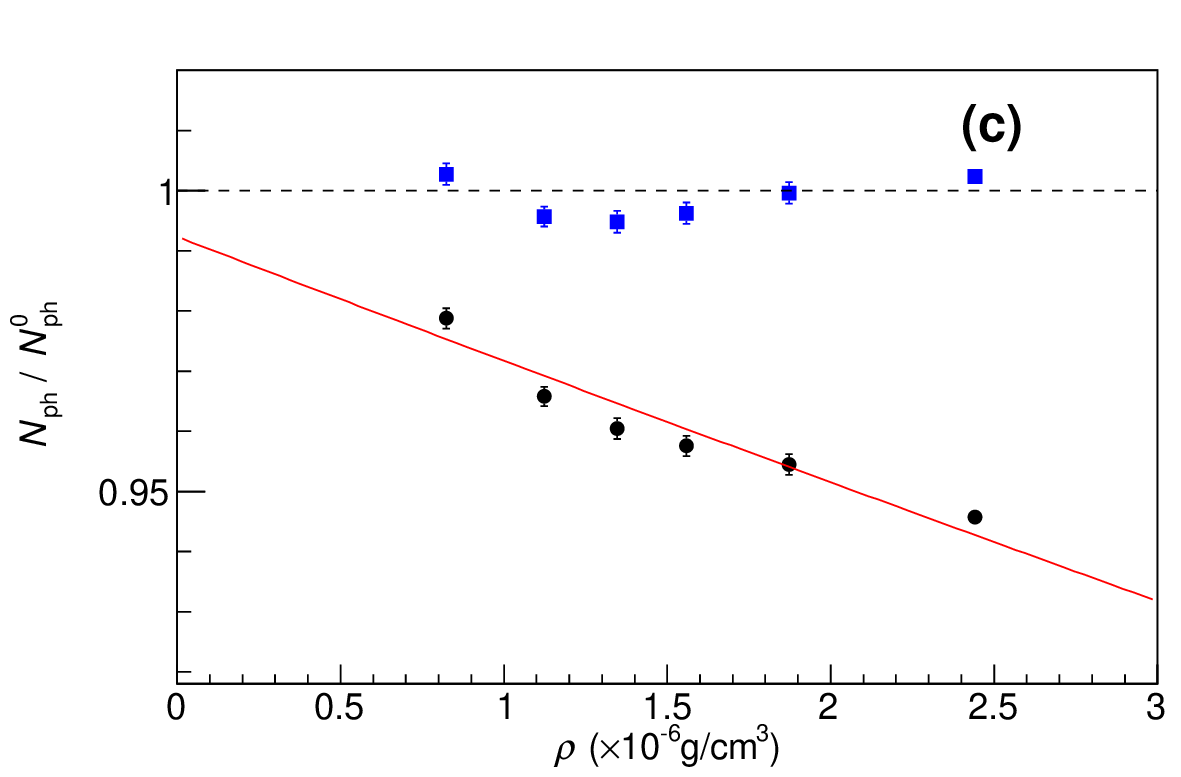}
    \caption{
    Following the CIC method, the relationship between muon content $N_\mu$ and atmospheric pressure was established, with the red line fitted using quadratic functions (a). The correlation between the number of secondary particles $N_\mathrm{e}$ and atmospheric pressure was also determined, with the red line fitted using quadratic functions (b). Additionally, $N_{\mathrm{ph}}$ in relation to absolute humidity was analyzed. The data (black dots) are obtained using the CIC method, while the expected values (red line) are derived from the simulation (c). 
    We corrected $N_\mu$ and $N_\mathrm{e}$ to an atmospheric pressure of 586.6 hPa, while $N_{\mathrm{ph}}$ was adjusted to an AOD of 0, aligning with the simulation settings. $N_\mu^{586.6}$ and $N_\mathrm{e}^{586.6}$ represent the muon content and secondary particles at 586.6 hPa, respectively, while $N_{\mathrm{ph}}^0$ represents the total number of photons at zero AOD. 
    The black dots represent the ratio of the measured signals from the CIC method to the expected signals at 586.6 hPa (or zero AOD), while the blue squares represent the corrected ratios based on the red lines.
    }
    \label{fig:CICCorrection}
\end{figure}

The measured energy spectrum ranges from 0.158 PeV to 12.6 PeV, within which the $X_{\mathrm{max}}$ of the proton-induced showers increases from 500 $\mathrm{g/cm^2}$ to 640 $\mathrm{g/cm^2}$.
The zenith angle of the data we used is around 45$^\circ$, corresponding to an atmospheric depth of 850 $\mathrm{g/cm^2}$.
Therefore, the events we measured have all developed beyond the maximum of the showers.
For these events, the measured secondary particles, including muons and electromagnetic particles, will vary with changes in atmospheric depth.
During the observation period, the atmospheric pressure variation at the LHAASO site was approximately 14 hPa.
Here, we use the constant intensity cut (CIC) method to correct the effects of atmospheric pressure changes on the data.
We divide the data into six parts based on the atmospheric pressure, with pressure ranges as [$<$591.0] hPa, [591.0,594.0] hPa, [594.0,595.5] hPa, [595.5,597.0] hPa, [597.0,598.5] hPa, and [$>$598.5] hPa.
The integral muon intensity of data across different atmospheric pressure ranges is shown in Fig.~\ref{fig:CICmethod}.
By cutting according to the flux intensity at $F(>N_\mu)=1.2\times 10^{-6}  \mathrm{m^{-2}sr^{-1}s^{-1}}$, the number of muons under different atmospheric pressures can be obtained.
The relationship between the number of muons obtained using CIC method and atmospheric pressure is shown in Fig.~\ref{fig:CICCorrection}a.
Using the same method, we can also obtain the relationship between $N_\mathrm{e}$ and atmospheric pressure (as shown in Fig.~\ref{fig:CICCorrection}b).
The following formula is used to fit the relationship between $N_\mu$, $N_\mathrm{e}$, and atmospheric pressure.
\begin{equation}
    N_{\mathrm{\mu/e}}=a_{\mathrm{\mu/e}}^0+a_{\mathrm{\mu/e}}^1\times P+a_{\mathrm{\mu/e}}^2\times P^2.
\end{equation}
Here, $a_{\mathrm{\mu/e}}^0$, $a_{\mathrm{\mu/e}}^1$, and $a_{\mathrm{\mu/e}}^2$ are the fitting parameters, and $P$ is the atmospheric pressure.
$N_\mu$ and $N_\mathrm{e}$ are the number of muons and the number of electromagnetic particles within the ring between 40 m and 200 m from the shower core, estimated
using the MDs and EDs located in that ring, respectively.
We correct $N_\mu$ and $N_\mathrm{e}$ to an atmospheric pressure of 586.6 hPa (consistent with the settings in the simulation).
After correction, the influence of atmospheric pressure on $N_\mu$ and $N_\mathrm{e}$ is shown in Fig.~\ref{fig:CICCorrection}a and Fig.~\ref{fig:CICCorrection}b as blue squares.

\subsubsection{Calibration of \texorpdfstring{$N_{\mathrm{ph}}$}{}}

The Cherenkov photons generated by the secondary charged particles in the shower are focused onto the surface of the SiPM camera by the spherical mirror of the WFCTA telescopes, where they are measured and digitized by a flash analog-to-digital converter. The photon response of the SiPM camera is calibrated and monitored in real-time using light emitting diodes (LEDs) of five different wavelengths, with a calibration uncertainty of less than 2.6\%~\cite{WFCTA-LED-calibration}.
The Cherenkov photons will be attenuated during their propagation in the atmosphere.
The main component responsible for the absorption of Cherenkov light in the atmosphere is the water vapor.
The water vapor content in the atmosphere ranges from 0\% to 4\%.
At the LHAASO site, the Aerosol Optical Depth (AOD) is measured by a Solar Radiometer, which quantifies the light absorbed and scattered by aerosols in a vertical column of the atmosphere during the day.
By combining this data with temperature and relative humidity measured at a meteorological station located at LHAASO site, we calculated the absolute humidity, thereby establishing the relationship between AOD and absolute humidity.
Through simulations, we calculated the attenuation coefficients at different AODs and, using the relationship between AOD and absolute humidity, derived the relationship between the attenuation coefficient and absolute humidity (as shown by the red line in Fig.~\ref{fig:CICCorrection}c).
On the other hand, we divided the data into six batches based on the absolute humidity in the atmosphere.
They are [$<$1.00]$\times10^{-6}$ $\mathrm{g/cm^3}$, [1.00,1.24]$\times10^{-6}$ $\mathrm{g/cm^3}$, [1.24,1.44]$\times10^{-6}$ $\mathrm{g/cm^3}$, [1.44,1.69]$\times10^{-6}$ $\mathrm{g/cm^3}$, [1.69,2.09]$\times10^{-6}$ $\mathrm{g/cm^3}$ and [$>$2.09]$\times10^{-6}$ $\mathrm{g/cm^3}$.
Using the CIC method, the relationship between $N_{\mathrm{ph}}/N_{\mathrm{ph}}^{0}$ and absolute humidity is shown in Fig.~\ref{fig:CICCorrection}c.
The results show that the observed Cherenkov light attenuation is highly consistent with the results obtained from simulations, with a difference of less than 1\%.

$N_{\mathrm{ph}}$ is corrected to an AOD of 0 (consistent with the settings in the simulation).
After correction, the influence of absolute humidity on $N_{\mathrm{ph}}$ is shown in Fig.~\ref{fig:CICCorrection}c as blue squares.

\section{ Simulation}
The CORSIKA software package~(version 77420)~\cite{CORSIKA-PACKAGE} is used to simulate the cascade process in the atmosphere.
Rayleigh scattering of Cherenkov photons by the atmosphere has also been considered by applying the U.S. Standard Atmosphere model in the simulation.
The secondary charged particles and the atmospheric Cherenkov photons in EAS are separately input into the simulation software of KM2A~\cite{KM2A-Geant4} and WFCTA for detector simulation.
For EDs and MDs, a simulation toolkit has been developed based on the GEANT4 package~\cite{KM2A-Geant4,GEANT4}. The simulation of the WFCTA telescopes is mainly based on a ray tracing procedure following every single Cherenkov photon generated in the air shower simulation using CORSIKA.
In this research, the number of events, $3.2\times10^7$, $3.2\times10^7$, and $2.9\times10^7$, based on three hadronic models, QGSJETII-04~\cite{QGSJETII-04}, EPOS-LHC~\cite{EPOS-LHC}, and SIBYLL 2.3d~\cite{SIBYLL2.3d}, were generated to describe high energy  hadronic interactions, respectively.
The interactions of particles with energy below $8\times10^{-5}$ PeV  have been modeled by a MonteCarlo  program  using the FLUKA code~\cite{fluka}.
Five groups of primary particles are generated, namely, protons, helium, nitrogen~(C-N-O group), aluminum~(Mg-Al-Si group), and iron.
The energy distribution of the events ranges from 0.01 PeV to 40 PeV.
Simulated events are generated according to a power law spectrum with the spectral index of $-1$.
The core positions are distributed within a square area with a side length of 1000 m, centered at the telescope array. The core positions are indicated by the area enclosed by the black line in Fig.~\ref{fig:coreselect1}a.
The shower directions are evenly distributed in the range of zenith angles of 35$^\circ$ to 55$^\circ$ and azimuth angles of 0$^\circ$ to 360$^\circ$.
For the simulated events, we utilized the same reconstruction and analysis algorithms as the experimental data to obtain the corresponding variables.

\section{ Data selection}

The LHAASO data used in this analysis were collected from October 2021 to April 2022.
To obtain a reliable data set, some quality selections have been applied according to the data status.
For KM2A, the following selection criteria were used:
\begin{enumerate}[label=(\arabic*),leftmargin=1.5cm]
    \item The reconstructed core position is constrained within the red line area in Fig.~\ref{fig:coreselect1}a. The red line consists of two parts: inner and outer. 
    The inner red line, which is 50 m away from the edge of the ED detector, is used to remove events with core positions inside WCDA.
    The arc at the lower right corner of the outer red line is 50 m away from the edge of the KM2A array.
    The two straight lines at the top and left side are to ensure that the selected area is more than 30 m away from the edge of the simulated events.
    \item The number of triggered EDs is more than 20.
    \item The range of $R_\mathrm{p}$ is from 180 m to 310 m.
    \item The direction of the event is within 10 degrees from the center of the telescope's field of view.
\end{enumerate}
For WFCTA, the following conditions were used to select good Cherenkov image events:
\begin{enumerate}[label=(\arabic*),leftmargin=1.5cm]
    \item The angular distance between the centroid of the Cherenkov image and the edge of the camera in the azimuth and zenith direction is greater than 3$^\circ$, as indicated by the red part in Fig.~\ref{fig:coreselect1}b.
    \item After event cleaning, the Cherenkov image still contains more than five tubes.
\end{enumerate}
The observations of the telescope are influenced by background light and weather conditions, so the data are also selected based on environmental conditions.
\begin{enumerate}[label=(\arabic*),leftmargin=1.5cm]
    \item The moonlight is the primary source of background light. When the moon passes through the telescope's field of view, the background light is too strong, causing a significant impact on the gain of SiPMs. At this time, the data is not used, and the telescope ceases observation.
    Only data meeting two conditions are used: the angle between the moon and the telescope’s main axis is greater than 20$\circ$, and the night sky background measured by WFCTA~\cite{WFCTA-construction} is less than 200 ADC count.
    \item The LHAASO site has installed an infrared cloud instrument to detect the infrared brightness temperature~($T_\mathrm{b}$) in the sky. A lower $T_\mathrm{b}$ indicates better weather conditions and therefore weaker atmospheric extinction of Cherenkov light. We only use the data with $T_\mathrm{b}$ less than \textcolor{black}{$-65^\circ$C}.
\end{enumerate}

After selection, the core resolution and the angular resolution are better than 6.5 m and 0.4$^\circ$, respectively, at energies above 0.158 PeV.
The high precision of both core position and arrival direction assures that the resolution of $R_\mathrm{p}$ is better than \textcolor{black}{3.5} m.
SDP resolution is better than 1.7$^\circ$, as shown in Fig.~\ref{fig:geo-resolution}.

\begin{figure}[htpb]
\centering\includegraphics[width=0.53\linewidth]{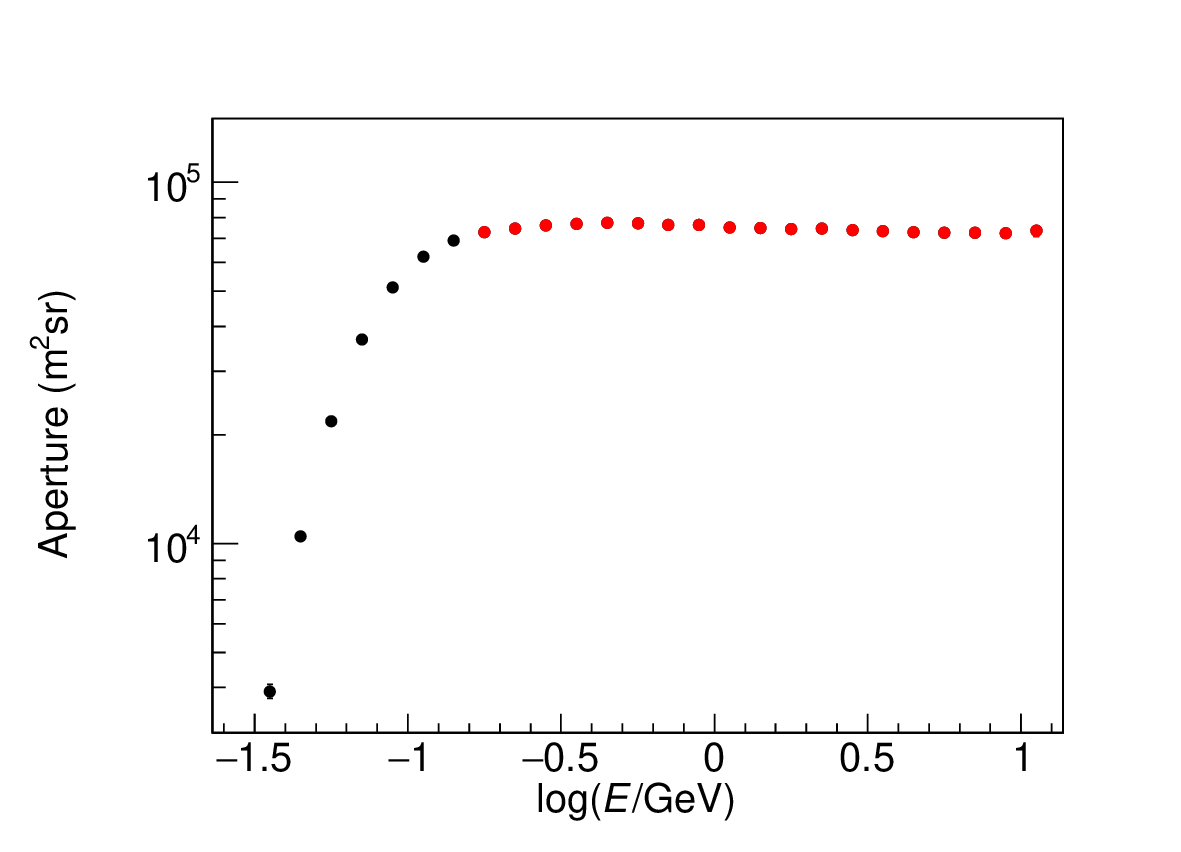}
\caption{The effective aperture of the KM2A and WFCTA hybrid measurement system after quality cuts is shown, with red dots marking the energy range of the proton energy spectrum discussed in this work.}
\label{fig:aperture}
\end{figure}

The aperture of the hybrid experiment for proton events is estimated with the aid of simulations as the fraction of the events \textcolor{black}{passing} all the KM2A and WFCTA selection cuts.
After applying the data quality criteria, the aperture gradually increases with energy and reaches a constant value above 0.2 PeV, which is approximately 75,000~m$^2$sr~(shown in Fig.~\ref{fig:aperture}).
The average observation time for each telescope in the hybrid observation experiment from October 2021 to April 2022 is approximately 900 hours.

\section{ Shower energy reconstruction}

For a well-developed air shower, all Cherenkov photons emitted by charged secondary particles (primarily positrons and electrons)
at all heights during the development of the air shower are accumulated when they touch down on the ground and form a so-called photon pool. The total number of photons in the pool is a good estimator of the energy of the entire shower.
Photons in the pool are distributed spatially following
the formula,
\begin{equation}
   \textcolor{black}{N_{\mathrm{ph}} = N_{\mathrm{ph}}^{250}\times \left(\frac{R_\mathrm{p}}{250}\right)^\alpha}.
    \label{fomular:NphNormal}
\end{equation}
We obtained the value of parameter $\alpha$ using proton events in the energy range of 0.2 PeV to 12.6 PeV, with the value of $\alpha=-2.35$.
A telescope, located at a distance $R_p$ from the shower, collects photons that fall into its spherical mirror collector. These photons are recorded by its SiPM camera to form a Cherenkov image.
We normalize the number of collected photons to 250 m from the shower center according to the lateral distribution, which is denoted as $N_{\mathrm{ph}}^{250}$. Then, the shower energy $E$ is measured.
A value of 250 m is selected as the median of the $R_\mathrm{p}$ distribution for all events used in the analysis.

\begin{figure}[htpb]
  \centering\includegraphics[width=0.92\linewidth]{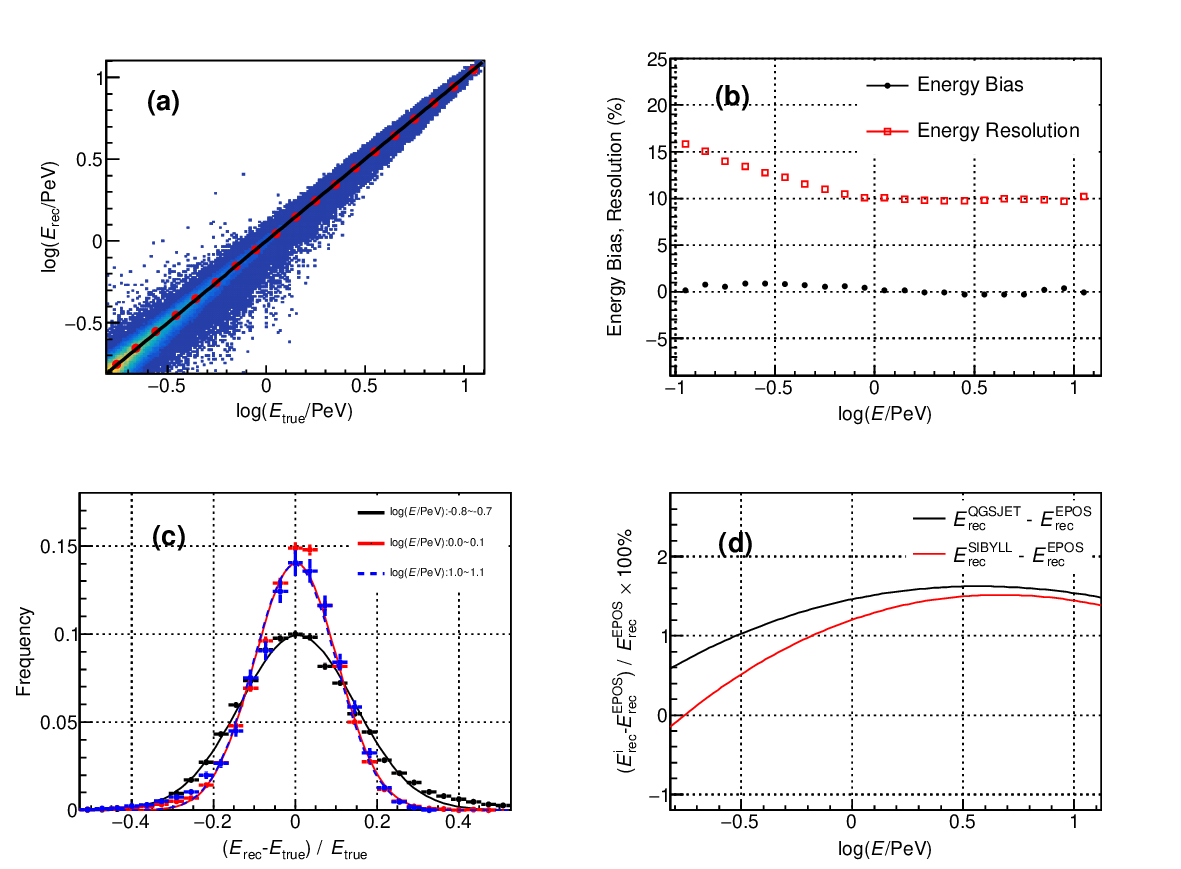}
    \caption{
 { The energy reconstruction of proton events. (a) The response function for proton events. The color scale indicates the  probability of simulated events within each bin. The red dots represent the mean values of the true energy and reconstructed energy. The black line indicates exact reconstruction of the energy.
 (b) The reconstructed energy bias is $<\pm2\%$~(black circles) and resolution~(red open squares) varies from 15\% to 10\% for proton events.
 (c) The energy resolution functions for events with reconstructed energy $\log(E_{\mathrm{rec}}/\mathrm{PeV})=-0.8$ to $-0.7$ (black), $\log(E_{\mathrm{rec}}/\mathrm{PeV})=0.0$ to $0.1$ (red), and $\log(E_{\mathrm{rec}}/\mathrm{PeV})=1.1$ to $1.2$ (blue). They are fitted by Gaussian functions.
 (d) The black solid curve represents the difference in reconstructed energy between the QGSJETII-04 model and the EPOS-LHC model. The red solid curve represents the difference in reconstructed energy between the SIBYLL 2.3d model and the EPOS-LHC model.
 }
}
  \label{fig:protonErecResult}
\end{figure}

Since muons carry energy from the hadronic components in the cascade process of a shower, the combination of  $N_{\mathrm{c\mu}}=N_{\mathrm{ph}}^{250}+a\times N_{\mu}$ can further improve the energy resolution, where $N_{\mu}$ is the number of muons  within the ring between 40 m and 200 m from the shower core, estimated using the MDs located in that ring.
Removing detectors within 40 m is intended to avoid the punch-through effect,
while removing detectors beyond 200 m is to prevent the introduction of significant noise.
The calculation method for $N_{\mu}$ can be found in \cite{WLP-Erec}.
From simulation studies, we obtained that the optimal
linear combination has the parameter $a = 3.0$.
Taking into account the very slight non-linear effect, the energy estimation,
\begin{equation}
\mathrm{log}(E/\mathrm{PeV}) = p_{0}+p_{1}\times \mathrm{log}(N_{\mathrm{c\mu}})+p_{2}\times \mathrm{log}^{2}(N_{\mathrm{c\mu}}),
\label{equation:Erec}
\end{equation}
results in a good energy response as shown in Fig~\ref{fig:protonErecResult}a, according to the full simulation of the detection.
Fig~\ref{fig:protonErecResult}b shows the Gaussian energy resolution functions for events in three energy bins.
Our energy reconstruction scheme ensures the energy resolution (the $\sigma$-parameter of Gaussian function) varies very little, ranging from 14\% at 0.158 PeV to 10\% at energies greater than 1 PeV.
The energy bias (the mean value of the Gaussian function) is less than $\pm2\%$ across the entire energy range for the pure proton event sample as shown in Fig~\ref{fig:protonErecResult}b.
We choose the bin-width for energy spectrum measurement as $\log(E/\mathrm{GeV})=0.1$.
The excellent properties, especially the symmetric Gaussian response functions, effectively minimize the bin-to-bin migration and avoid distortion in the reconstructed energy spectrum. This makes LHAASO the ideal detector for precision measurement of energy spectrum which is characterized by some systematic structures on top of a single index power-law functional form.
The fit parameters $p_{0}$, $p_{1}$ and $p_{2}$ for QGSJETII-04, EPOS-LHC and SIBYLL 2.3d are listed in Table~\ref{table:Erecparameter}.
In the simulation, the ratios of electromagnetic components and hadronic components in an EAS vary according to  the specific hadronic interaction models used. Our energy reconstruction scheme measures both components simultaneously, thereby minimizing the discrepancies in reconstructed energies arising from hadronic interaction models.
The differences in reconstructed energy obtained from various models are less than 1.7\%, as shown in Fig.~\ref{fig:protonErecResult}d, and values at three energies are listed in the fifth column of Table~\ref{table:Erecparameter} as representatives.
The last column of Table~\ref{table:Erecparameter} shows the differences in energy resolution obtained from three models, which are within 1\%.

\begin{table}[htbp]
\renewcommand{\arraystretch}{1.25}
\begin{center}
\begin{threeparttable}
\caption{The fitting results of each parameter in Eq.~(\ref{equation:Erec}) for different hadronic interaction models.}
\begin{tabular}{cccccccc}
\hline
\multirow{2}{*}{Model} & \multirow{2}{*}{$p_{0}$} & \multirow{2}{*}{$p_{1}$} & \multirow{2}{*}{$p_{2}$} & \multicolumn{3}{c}{$(E_{\mathrm{rec}}-E_{\mathrm{rec}}^{\mathrm{EPOS}})/E_{\mathrm{rec}}^{\mathrm{EPOS}}$} &\multirow{2}{*}{$(E_{\mathrm{Res}}-E_{\mathrm{Res}}^{\mathrm{EPOS}})/E_{\mathrm{Res}}^{\mathrm{EPOS}}$}\\
 & & & & 0.316 PeV & 1 PeV & 3.16 PeV & \\
\hline
 QGSJETII-04 & -3.519 & 0.6376 & 0.03536 & 1.0\% & 1.5\% & 1.6\% & 0.8\% \\
 EPOS-LHC & -3.475 & 0.6173 & 0.03738 & 0 & 0 & 0 & 0 \\
 SIBYLL 2.3d & -3.544 & 0.6468 & 0.03450 & 0.5\% & 1.2\% & 1.5\% & 0.5\% \\
\hline
\end{tabular}
\label{table:Erecparameter}
\begin{tablenotes}
\item The fifth column shows the difference in the reconstructed energy obtained using different model parameters at various energy compared to that obtained using EPOS-LHC model parameters.
The last column shows the maximum difference in energy resolution between the results obtained from different models and the EPOS-LHC model.
\end{tablenotes}
\end{threeparttable}
\end{center}
\end{table}

\section{Proton selection~\label{section:protonSelection}}
\subsection{Component sensitive parameters}

\begin{figure}[htpb]
\centering\includegraphics[width=0.95\linewidth]{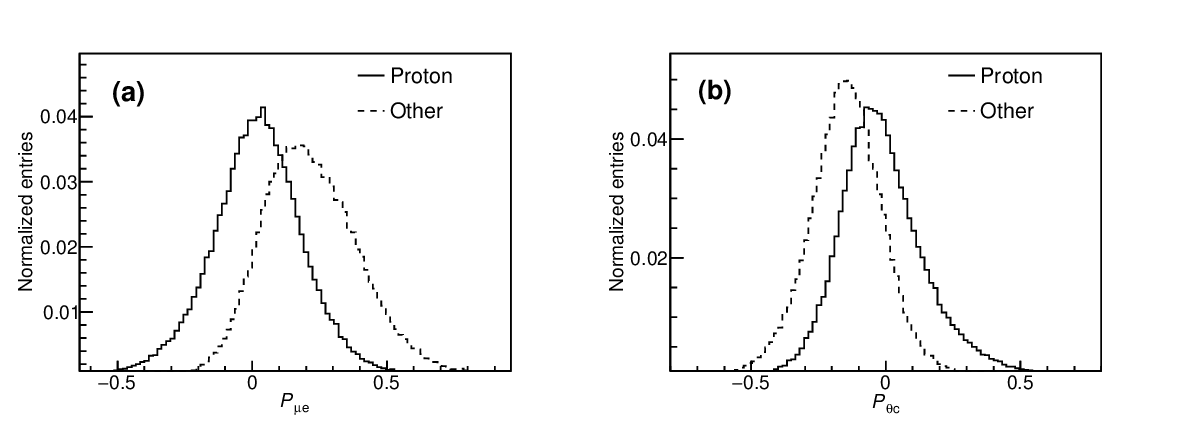}
\caption{Distribution of component sensitive parameter. (a): The $P_{\mathrm{\mu e}}$ distributions for protons (solid line) and other CR particles (dashed line) in the energy range of 0.158 PeV to 12.6 PeV, both normalized to an integral of 1. (b): the distribution of $P_{\mathrm{\theta c}}$ for protons~(solid line) and other primaries~(dashed line) with energy between 0.158 PeV and 12.6 PeV. Both distributions are normalized to have an integral of 1.}
\label{figure:component-Pthetac}
\end{figure}

Electromagnetic and muonic components in EASs are related to the primary energy and atomic number~($A$).
For primary particles with the same energy, the larger the $A$, the more muons there are in the shower.
The relationship among muons, electrons, primary energy, and $A$ is expressed as
\begin{equation}
    N_{\mu}=K_\mu A^{1-\beta}\left(\frac{E_0}{1~\rm{PeV}}\right)^\beta,
    \label{fomula:NmuA}
\end{equation}
\begin{equation}
    N_{\mathrm{e}}=K_\mathrm{e} A^{1-\alpha}\left(\frac{E_0}{1~\rm{PeV}}\right)^\alpha,
    \label{fomula:NeA}
\end{equation}
the values of $\alpha$ and $\beta$ are related to the interaction model~\cite{EASmodel}.
Solving for $A$ one obtains:
\begin{equation}
    \log A=\frac{\alpha}{\alpha-\beta}\log \left(\frac{N_\mu}{N_\mathrm{e}^{\beta/\alpha}}\right)-\frac{\alpha}{\alpha-\beta}\log \left(\frac{K_\mu}{K_\mathrm{e}^{\beta/\alpha}}\right)=\frac{\alpha}{\alpha-\beta}\log\left(\frac{N_\mu}{N_\mathrm{e}^{\beta/\alpha}}\right)+\mathrm{const}.
\end{equation}
This suggests introducing the mass sensitive parameter $P_{\mathrm{\mu e}}$
\begin{equation}
    P_{\mathrm{\mu e}}=\mathrm{log}\frac{N_{\mu}}{N_{\mathrm{e}}^{\beta/\alpha}}+1.76.
\end{equation}
This parameter is linearly related to $\log A$.
The average value of $\log\frac{N_{\mu}}{N_{\mathrm{e}}^{\beta/\alpha}}$ for all proton events in the reconstructed energy range from 0.158 PeV to 12.6 PeV is $-1.76$.
It is important to note that since $\alpha>1$ and $\beta<1$, $N_{\mu}$ is positively correlated with $A$, while $N_{\mathrm{e}}$ is negatively correlated with $A$.
Using the ratio $N_{\mu}/N_{\mathrm{e}}^{\beta/\alpha}$ not only eliminates the energy dependence but also enhances the component discrimination capability compared to using $N_{\mu}$ alone.
In this work, $N_{\mu}$ and $N_{\mathrm{e}}$ are muons and electromagnetic particles within a range of 40 m to 200 m from the shower axis.
Based on the simulation results, in order to eliminate the dependence of variable $P_{\mathrm{\mu e}}$ on energy, $\beta/\alpha=0.82$ is used.
Fig.~\ref{figure:component-Pthetac}a shows the distribution of the parameter $P_{\mathrm{\mu e}}$ for protons and other components.

The larger the mass number $A$ of the primary particle, the larger the cross section for
interactions with atomic nuclei in the atmosphere is,
resulting in a shallower average atmospheric depth for the first interaction.
According to the "superposition model", for primary particles of energy $E$ and mass
number $A$, each nucleon carries the energy $E_0 = E/A$, and therefore the showers of large
mass nuclei are less penetrating and develop higher in the atmosphere than protons of the
same energy.
At the same energy, the $X_{\mathrm{max}}$ of protons is about 100 g/cm$^2$ larger than that of iron nuclei~\cite{EASmodel-Matthews}.
The long axis of the Cherenkov image corresponds to the longitudinal development direction of the EAS.
The number of secondary particles and Cherenkov photons in the EAS increases as the shower develops, and gradually decreases after reaching the maximum.
On the Cherenkov image, the SiPM signal first increases along the long axis, and then gradually decreases.
Therefore, the angular distance, measured in degrees, from the centroid of the Cherenkov image to the arrival direction of the event~($\theta_{\mathrm{c}}$) is related to $X_{\mathrm{max}}$.

The parameter $\theta_{\mathrm{c}}$ is also related to the primary energy and geometry. Therefore we normalized $\theta_{\mathrm{c}}$ with the reconstructed $R_\mathrm{p}$ and reconstructed shower direction. The formula is
\begin{equation}
    \theta_{\mathrm{c}}^{250} = \frac{\theta_{\mathrm{c}}}{\cos\theta}-0.011\times (R_\mathrm{p}-250).
\end{equation}
Here, $\theta$ is the zenith angle of the event. We use $\theta_{\mathrm{c}}^{250}$ to build the composition-sensitive parameter,
\begin{equation}
P_{\theta {\mathrm{c}}}=\frac{\theta_{\mathrm{c}}^{250}-\langle \theta_{\mathrm{c}}^{250}|_\mathrm{p}(E)\rangle}{\langle \theta_{\mathrm{c}}^{250}|_\mathrm{p}(E=1PeV)\rangle},
\end{equation}
where $\langle \theta_{\mathrm{c}}^{250}|_\mathrm{p}(E)\rangle$ is the mean value of $\theta_{\mathrm{c}}^{250}$ for proton events.
The following formula is used to fit the relationship between $\langle \theta_{\mathrm{c}}^{250}|_\mathrm{p}(E)\rangle$ and primary energy.
\begin{equation}
    \langle \theta_{\mathrm{c}}^{250}|_\mathrm{p}(E)\rangle = p0+p1\times \mathrm{log}E+p2\times \mathrm{log}^2E.
\end{equation}
The fit result is $p0 = -5.492$, $p1 = 2.548$, $p2 = -0.153$ and $\langle \theta_{\mathrm{c}}^{250}|_\mathrm{p}(E=1\mathrm{PeV})\rangle=4.29$.
Fig.~\ref{figure:component-Pthetac}b shows the distribution of the parameter $P_{\mathrm{\theta {c}}}$ for protons and other components.

\subsection{Selection of proton events}

\begin{figure}[htpb]
\centering\includegraphics[width=0.99\linewidth]{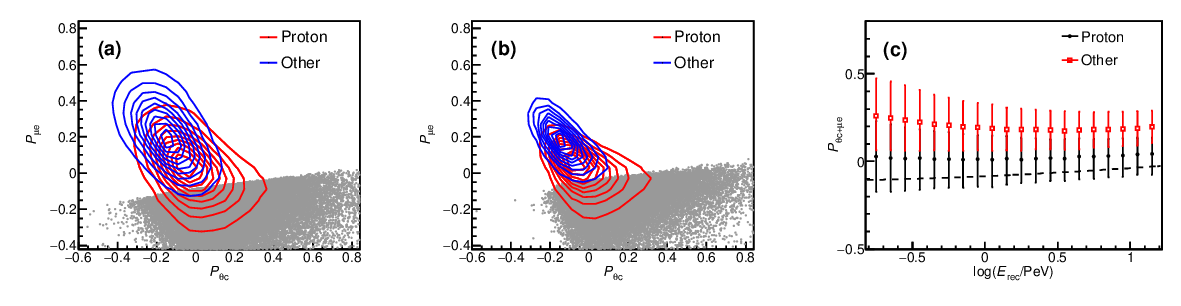}
\caption{Proton selection. (a): \textcolor{black}{The correlation plot illustrates $P_{\mathrm{\theta c}}$ versus $P_{\mathrm{\mu e}}$  for proton (red contour) and other CR components (blue contour) within the energy range of 0.158 PeV to 1 PeV. The gray dots represent the distributions of $P_{\mathrm{\theta c}}$ and $P_{\mathrm{\mu e}}$ for events after applying the proton selection criteria.}
(b): The correlation plot illustrates $P_{\mathrm{\theta c}}$ versus $P_{\mathrm{\mu e}}$  for proton (red contour) and other CR components (blue contour) within the energy range of 1 PeV to 12.6 PeV. The gray dots represent the distributions of $P_{\mathrm{\theta c}}$ and $P_{\mathrm{\mu e}}$ for events after applying the proton selection criteria.
(c): The variation of $P_{\mathrm{\theta c+\mu e}}$ for proton and other CR components over energy. The error bars represent the standard deviation of $P_{\mathrm{\theta c+\mu e}}$ of different energy bins. The black dashed line indicates the criteria used to select proton events from the all CR particles data.
}
\label{figure:select_result1}
\end{figure}

\begin{figure}[htpb]
    \centering
    \includegraphics[width=0.95\linewidth]{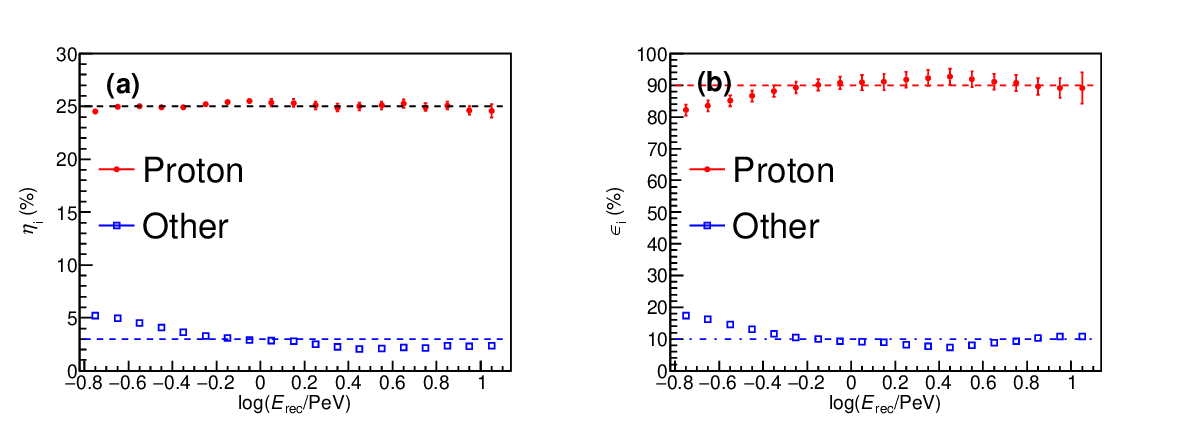}
    \caption{
   (a) After applying the selection criteria, the relationship between the proton selection efficiency and reconstruction energy (red dot), as well as the relationship between the selection efficiency of other components and reconstruction energy (blue square), are shown. The two dashed lines at 25\% (black) and 3\%(blue) serve as their reference values, respectively. The selection efficiency of protons is 8.5 times higher than that of the other components at 1 PeV, indicating that the suppression ratio for the other components is 8.5 times. This suppression ratio increases with energy, from 5 times at 0.158 PeV to 10 times at 2 PeV, and then remains unchanged as the energy increases.
   (b) Purity refers to the ratio of retained proton samples to all retained samples, and its relationship with energy is shown (red dot). Contamination refers to the ratio of retained other component or non-proton samples to all retained samples, and its relationship with energy is also shown (blue square). The two dashed lines at 90\% (red) and 10\%(blue) serve as their reference values, respectively.}
    \label{figure:select_result_puritySelection}
\end{figure}

In this paper, proton events are selected by combining the parameters $P_{\mathrm{\theta {c}}}$ and $P_{\mathrm{\mu e}}$.
Fig.~\ref{figure:select_result1}a \textcolor{black}{and Fig.~\ref{figure:select_result1}b} show the correlation of $P_{\mathrm{\theta c}}$ and $P_{\mathrm{\mu e}}$ for proton events and other events.
We combine $P_{\mathrm{\theta {c}}}$ and $P_{\mathrm{\mu e}}$ according to the following formula,
\begin{equation}
    P_{\mathrm{\theta c+\mu e}} = -\sin(\delta)\cdot P_{\mathrm{\theta c}}+\cos(\delta)\cdot P_{\mathrm{\mu e}}.
\end{equation}
To optimize the purity of the protons, the value of $\delta=8.5^\circ$ is adopted.
Fig.~\ref{figure:select_result1}c shows the variation of $P_{\mathrm{\theta c+\mu e}}$ with energy.
By balancing the small contamination from helium and the high survival proportion of proton events, we determined the dashed black line in the figure to select proton events.
The selection criteria for protons are as follows,
\begin{equation}
     P_{\mathrm{\theta c+\mu e}}<P2\cdot \mathrm{log}^2E+P1\cdot \mathrm{log}E+P0,
    \label{fomula:protonSelect}
\end{equation}
where $P0 = -0.084$, $P1 = 0.036$, and $P2 = 0.01$.
Fig.~\ref{figure:select_result_puritySelection} shows the selection efficiency and purity results at different energies.
The definitions of purity and selection efficiency are as follows,
\begin{equation}
    \epsilon_\mathrm{P} = \frac{N_{\mathrm{s}}^{\mathrm{P}}}{N_{\mathrm{s}}^{\mathrm{P}}+N_{\mathrm{s}}^{\mathrm{H}}},
    \label{fomula:ProtonPurity}
\end{equation}
\begin{equation}
    \eta_\mathrm{P} = \frac{N_{\mathrm{s}}^{\mathrm{P}}}{N_{0}^{\mathrm{P}}},
    \label{fomula:ProtonSelect}
\end{equation}
where $N_{\mathrm{s}}^{\mathrm{P}}$, $N_{\mathrm{s}}^{\mathrm{H}}$, and $N_{0}^{\mathrm{P}}$ are the numbers of selected protons, selected heavier particles, and total protons before selection, respectively.

When events are cut according to the selection criteria of Eq.~(\ref{fomula:protonSelect}), the C-N-O and heavier components are nearly completely suppressed, with a survival probability of less than 0.02\%. This leads to a contamination less than 0.5\% in the proton sample throughout the entire energy range. It is then negligible.
In the proton sample, the average contamination from other particles, which is almost helium, is 10.7\%, as indicated by the blue squares in Fig.~\ref{figure:select_result_puritySelection}b.
It is $\lesssim11\%$ for energies above 0.5 PeV and $>$15\% only in two energy bins below 0.25 PeV.
The high purity that we achieved above 0.5 PeV is already compatible with what was obtained in space-borne experiments in the high-energy range.
For instance, the fraction of helium contamination in the proton sample for DAMPE is $\sim$5\% around 50 TeV~\cite{DAMPE:2019gys}.
This is an achievement that was never reached in ground-based experiments.

\section{Proton energy spectrum}
\begin{figure}[htpb]
\centering\includegraphics[width=0.55\linewidth]{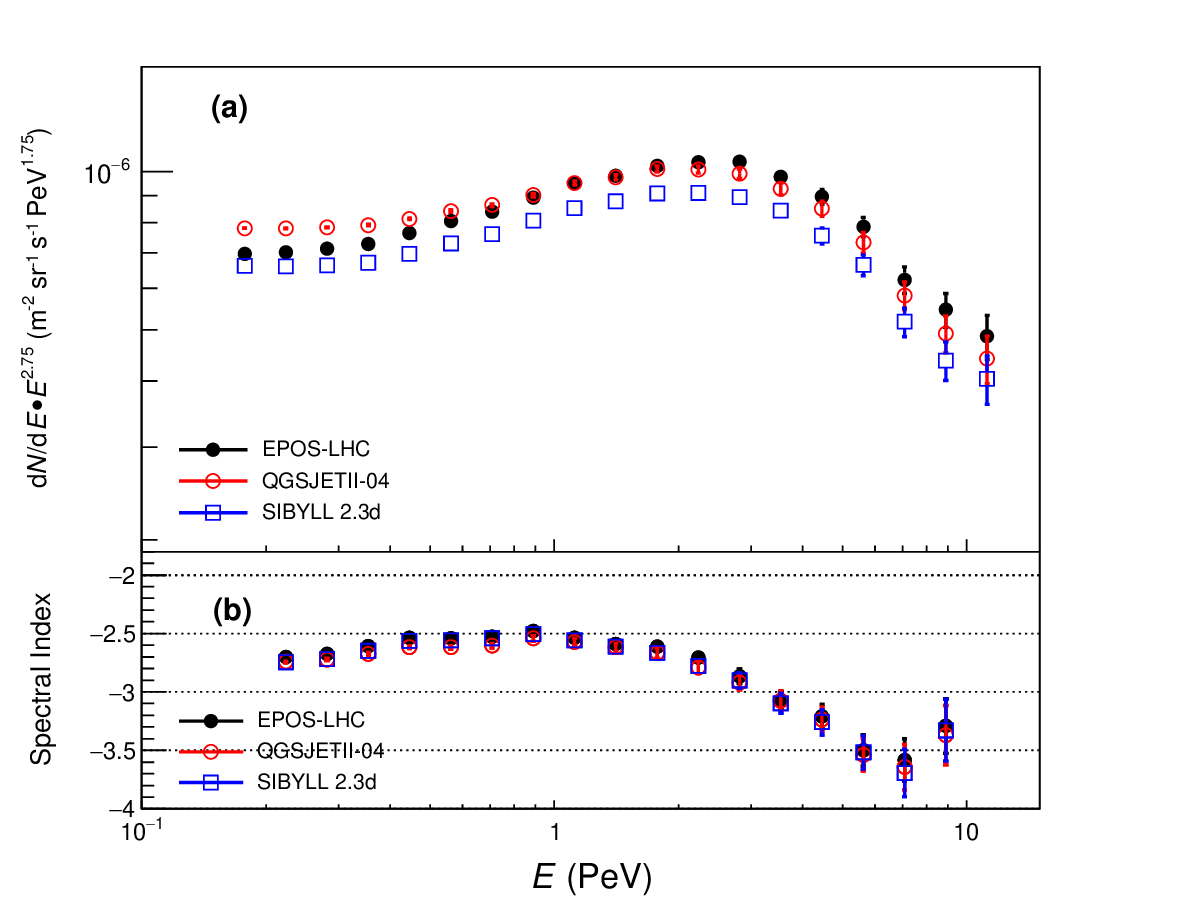}
\caption{ The energy spectrum of CR protons measured by LHAASO using different hadronic interaction models: EPOS-LHC (black dot), QGSJETII-04 (red circle), and SIBYLL 2.3d (blue square). (a) Proton flux multiplied by $E^{2.75}$ as a function of energy, with error bars representing statistical uncertainties. (b) The local spectral index as a function of energy, derived from fitting three adjacent data points with a power-law function. This shows a slight hardening of $\Delta\gamma\sim0.2$  and a gradual softening (``knee'') with $\Delta\gamma\sim-1$. Different hadronic interaction models show the same feature.}
\label{fig:protonSpectrum-Fit}
\end{figure}

Fig.~\ref{fig:protonSpectrum-Fit} presents the proton energy spectrum obtained based on the QGSJETII-04, EPOS-LHC and SIBYLL 2.3d models.
The values of the proton energy spectrum from the LHAASO experiment, along with the corresponding statistical uncertainty and combined systematic uncertainties (excluding the uncertainty from hadronic interaction models), are listed in Table~\ref{table:protonSpectrumText}.

\begin{table}[htbp]
\renewcommand{\arraystretch}{1.25}
\begin{center}
\begin{threeparttable}
\caption{Proton energy spectrum flux measured by LHAASO.}
\setlength{\tabcolsep}{3pt}
\begin{tabular}{clcccc}
\hline
Energy & Number of events & Flux$\pm$stat.$\pm$syst. (QGSJETII-04) & Flux$\pm$stat.$\pm$syst. (EPOS-LHC) & Flux$\pm$stat.$\pm$syst. (SIBYLL2.3d)\\
$\mathrm{log(E/PeV)}$ & & $\mathrm{PeV^{-1}}$ $\mathrm{m^{-2}}$ $\mathrm{s^{-1}}$ $\mathrm{sr^{-1}}$ & $\mathrm{PeV^{-1}}$ $\mathrm{m^{-2}}$ $\mathrm{s^{-1}}$ $\mathrm{sr^{-1}}$ & $\mathrm{PeV^{-1}}$ $\mathrm{m^{-2}}$ $\mathrm{s^{-1}}$ $\mathrm{sr^{-1}}$\\
\hline
 $-0.8$ -- $-0.7$ & 236775 & (9.051$\pm$0.019$\pm$0.513)$\times 10^{-5}$ &  (8.084$\pm$0.017$\pm$0.458)$\times 10^{-5}$  &  (7.667$\pm$0.016$\pm$0.434)$\times 10^{-5}$ \\
 $-0.7$ -- $-0.6$ & 160787 & (4.803$\pm$0.012$\pm$0.274)$\times 10^{-5}$ &  (4.326$\pm$0.011$\pm$0.247)$\times 10^{-5}$  &  (4.061$\pm$0.010$\pm$0.232)$\times 10^{-5}$ \\
 $-0.6$ -- $-0.5$ & 108986 & (2.561$\pm$0.008$\pm$0.147)$\times 10^{-5}$ &  (2.331$\pm$0.007$\pm$0.134)$\times 10^{-5}$  &  (2.166$\pm$0.007$\pm$0.124)$\times 10^{-5}$ \\
 $-0.5$ -- $-0.4$ & 73957  & (1.372$\pm$0.005$\pm$0.080)$\times 10^{-5}$ &  (1.264$\pm$0.005$\pm$0.073)$\times 10^{-5}$  &  (1.163$\pm$0.004$\pm$0.067)$\times 10^{-5}$ \\
 $-0.4$ -- $-0.3$ & 51111  & (7.491$\pm$0.033$\pm$0.440)$\times 10^{-6}$ &  (7.030$\pm$0.031$\pm$0.413)$\times 10^{-6}$  &  (6.426$\pm$0.028$\pm$0.377)$\times 10^{-6}$ \\
 $-0.3$ -- $-0.2$ & 35405  & (4.114$\pm$0.022$\pm$0.245)$\times 10^{-6}$ &  (3.932$\pm$0.021$\pm$0.235)$\times 10^{-6}$  &  (3.573$\pm$0.019$\pm$0.213)$\times 10^{-6}$ \\
 $-0.2$ -- $-0.1$ & 24280  & (2.241$\pm$0.014$\pm$0.136)$\times 10^{-6}$ &  (2.177$\pm$0.014$\pm$0.132)$\times 10^{-6}$  &  (1.976$\pm$0.013$\pm$0.120)$\times 10^{-6}$ \\
 $-0.1$ -- $0.0$ & 17128   & (1.243$\pm$0.009$\pm$0.077)$\times 10^{-6}$ &  (1.230$\pm$0.009$\pm$0.077)$\times 10^{-6}$  &  (1.111$\pm$0.008$\pm$0.069)$\times 10^{-6}$ \\
 $0.0$ -- $0.1$ & 11987    & (6.955$\pm$0.064$\pm$0.446)$\times 10^{-7}$ &  (6.954$\pm$0.064$\pm$0.446)$\times 10^{-7}$  &  (6.237$\pm$0.057$\pm$0.400)$\times 10^{-7}$ \\
 $0.1$ -- $0.2$ & 8220     & (3.789$\pm$0.042$\pm$0.252)$\times 10^{-7}$ &  (3.811$\pm$0.042$\pm$0.254)$\times 10^{-7}$  &  (3.411$\pm$0.038$\pm$0.227)$\times 10^{-7}$ \\
 $0.2$ -- $0.3$ & 5652     & (2.085$\pm$0.028$\pm$0.145)$\times 10^{-7}$ &  (2.113$\pm$0.028$\pm$0.147)$\times 10^{-7}$  &  (1.873$\pm$0.025$\pm$0.130)$\times 10^{-7}$ \\
 $0.3$ -- $0.4$ & 3821     & (1.104$\pm$0.018$\pm$0.081)$\times 10^{-7}$ &  (1.140$\pm$0.018$\pm$0.084)$\times 10^{-7}$  &  (9.961$\pm$0.161$\pm$0.732)$\times 10^{-8}$ \\
 $0.4$ -- $0.5$ & 2527     & (5.752$\pm$0.114$\pm$0.452)$\times 10^{-8}$ &  (6.064$\pm$0.121$\pm$0.476)$\times 10^{-8}$  &  (5.195$\pm$0.103$\pm$0.408)$\times 10^{-8}$ \\
 $0.5$ -- $0.6$ & 1581     & (2.857$\pm$0.072$\pm$0.243)$\times 10^{-8}$ &  (3.014$\pm$0.076$\pm$0.256)$\times 10^{-8}$  &  (2.600$\pm$0.065$\pm$0.221)$\times 10^{-8}$ \\
 $0.6$ -- $0.7$ & 971      & (1.39$\pm$0.05$\pm$0.13)$\times 10^{-8}$    &  (1.47$\pm$0.05$\pm$0.14)$\times 10^{-8}$     &  (1.24$\pm$0.04$\pm$0.12)$\times 10^{-8}$ \\
 $0.7$ -- $0.8$ & 571      & (6.38$\pm$0.27$\pm$0.66)$\times 10^{-9}$    &  (6.83$\pm$0.29$\pm$0.71)$\times 10^{-9}$     &  (5.78$\pm$0.24$\pm$0.60)$\times 10^{-9}$ \\
 $0.8$ -- $0.9$ & 306      & (2.68$\pm$0.15$\pm$0.31)$\times 10^{-9}$    &  (2.87$\pm$0.16$\pm$0.34)$\times 10^{-9}$     &  (2.39$\pm$0.14$\pm$0.28)$\times 10^{-9}$ \\
 $0.9$ -- $1.0$ & 180      & (1.21$\pm$0.09$\pm$0.16)$\times 10^{-9}$    &  (1.34$\pm$0.10$\pm$0.18)$\times 10^{-9}$     &  (1.07$\pm$0.08$\pm$0.14)$\times 10^{-9}$ \\
 $1.0$ -- $1.1$ & 109      & (5.74$\pm$0.55$\pm$0.90)$\times 10^{-10}$    &  (6.33$\pm$0.61$\pm$0.99)$\times 10^{-10}$     &  (5.25$\pm$0.50$\pm$0.82)$\times 10^{-10}$ \\
\hline
\end{tabular}
\label{table:protonSpectrumText}
\begin{tablenotes}
\item The first column corresponds to the energy intervals of different data points. The second column represents the number of remaining events after proton selection within these energy intervals. The last three columns represent the proton fluxes obtained under different hadron interaction models, together with statistical and systematic uncertainties. The systematic uncertainties for each item are listed in Table~\ref{table:systematicUncertainty}.
\end{tablenotes}
\end{threeparttable}
\end{center}
\end{table}

\section{Systematic uncertainty analysis}

We considered the following aspects of systematic uncertainties in both the reported flux and the energy measurement of shower events.

\subsection{Uncertainties in the flux evaluation}

\subsubsection{Hadronic interaction models}

With regard to the uncertainties from the hadronic models, we generated events using three different hadronic interaction models: QGSJETII-04, EPOS-LHC, and SIBYLL 2.3d. Each model features different interaction cross-sections, multiplicities, and production ratios of hadrons, ultimately leading to variations in the observables of EAS. 
For instance, the interaction cross-section of particles influences the step length between each interaction, leading to differences in the $X_{\mathrm{max}}$. Additionally, the decay of hadrons produces muons, and varying production ratios of hadrons within the models will result in different muon content in the shower.
Using these interaction models, we re-analyzed the whole data set following the same analysis process as what was done to obtain the proton energy spectrum described above. Figure~\ref{fig:protonSpectrum-Fit} illustrates the proton energy spectrum obtained from the three interaction models. The largest difference in the proton energy spectra between these models is approximately 17\%.

\subsubsection{Composition models}

\begin{figure}[htpb]
\centering\includegraphics[width=0.55\linewidth]{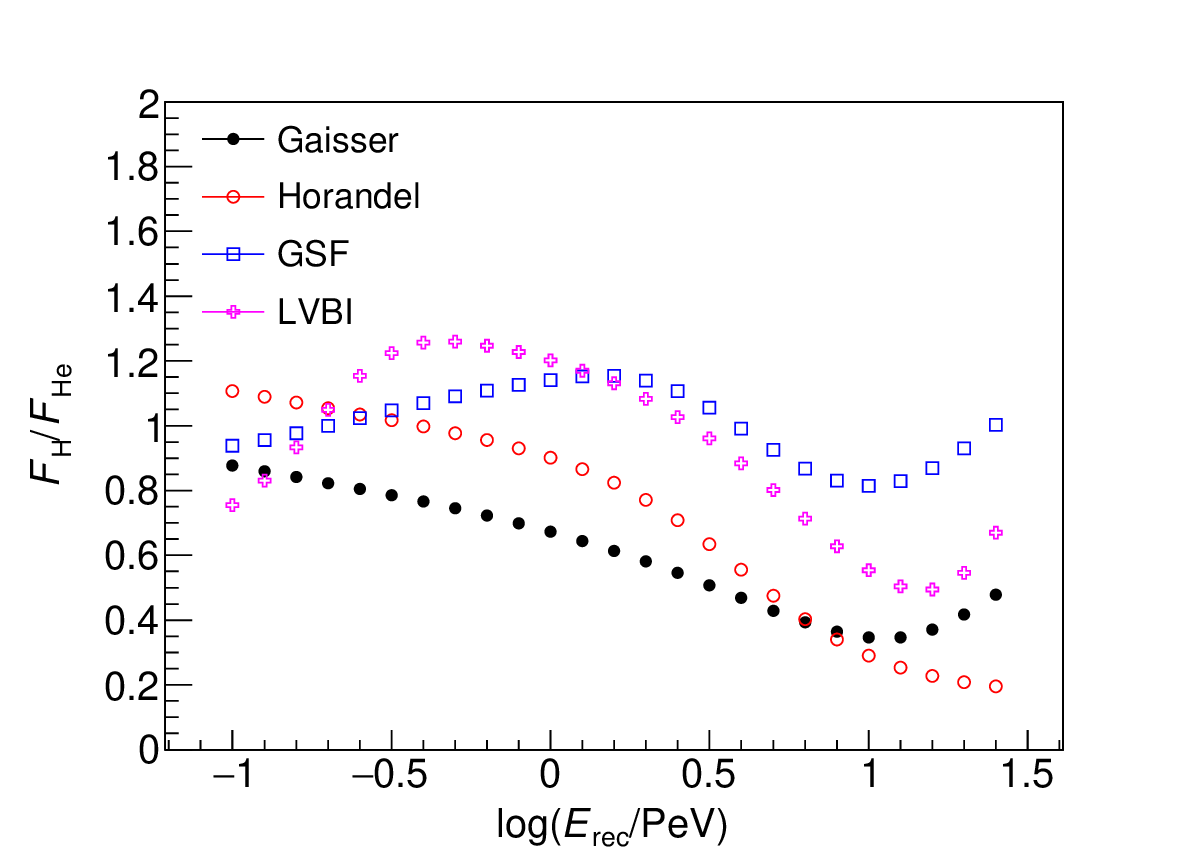}
\caption{The $H/He$ as a function of energy for different component models (GSF, Gaisser, Horandel, and LVBI) shows that the relative difference between these models increases from about 1.46 times at 0.1 PeV to about 2.7 times at 10 PeV.}
\label{fig:pHeRatio}
\end{figure}

According to the results in Sec.~\ref{section:protonSelection}, the main contamination in proton samples comes from helium nuclei. After proton selection, the contamination of CNO and heavier components is less than 0.5\%.
The proton selection efficiency at 1 PeV is approximately 8.5 times higher than that of other particles~(almost helium), which means the change in purity $\epsilon$ is 1/8.5 times that of the change in the ratio of protons to helium fluxes ($H/He$).
In this work, we use four composition models to study the impact of the different $H/He$ on the energy spectrum.
These models are Global Spline Fit~(GSF) model~\cite{GSFmodel}, Gaisser model~\cite{Gaisser:2013bla}, Horandel model~\cite{Horandelmodel} and LVBI model~\cite{BXJmodel}.
The higher the proportion of protons relative to helium nuclei in the primary CRs, the higher the purity $\epsilon_{\mathrm{P}}$ of the proton samples obtained after selection in our analysis.
The $H/He$ of the four composition models is shown in Fig.~\ref{fig:pHeRatio}.

\begin{figure}[htpb]
\centering\includegraphics[width=0.55\linewidth]{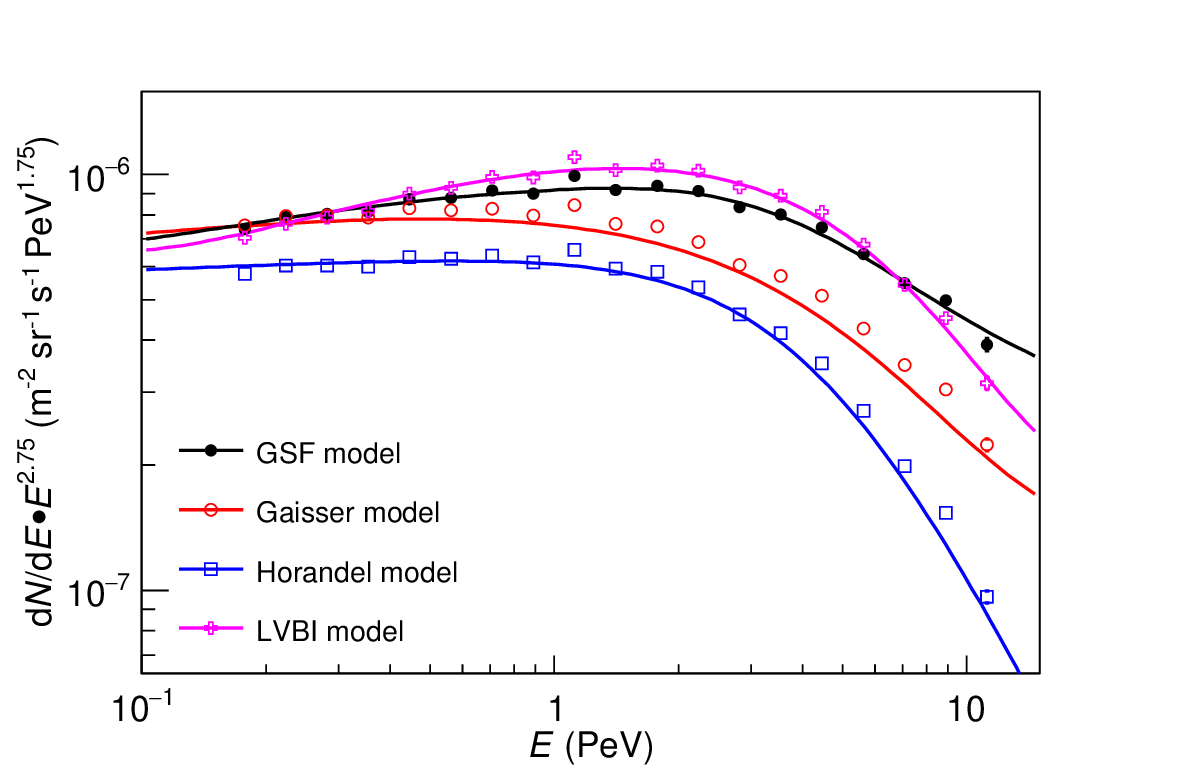}
\caption{ The energy spectra analysis process is established using the GSF composition model and validated with simulation data from different component models: GSF, Gaisser, Horandel, and LVBI. The lines show expected spectra of each component model, while circles indicate reconstructed results with discrepancies of 3\% -- 5\% for energies below 1 PeV, about 7\% for 3 PeV and 15\% for 10 PeV.
}
\label{figure:systematic1}
\end{figure}

We use the GSF model to optimize the proton selection criteria. Applying these criteria to the simulated data set with the specific mixture of 5 species, we obtain a proton sample with the  average purity $\bar{\epsilon_\mathrm{p}}\sim$ 89.3\%.
The survival probability (i.e. efficiency $\eta_\mathrm{p}$) of proton events is 25\%. 
Subsequently, these selection criteria are applied to data sets generated using different composition model assumptions, thereby yielding different purities of protons. Finally, the energy spectra of the various composition models are reconstructed, with the results illustrated by the points in Fig.~\ref{figure:systematic1}.
The spectrum of the GSF model is perfectly reconstructed within the statistical errors.
This test validates the robustness of our analysis method. At first, the selection of 25\% of events yields an unbiased sample that allows to reconstruct the correct energy spectrum.
Secondly, applying the criteria to the simulated data with other models, which are quite different from each other in terms of $H/He$,   as shown in Fig.~\ref{fig:pHeRatio}, all the spectra are reconstructed with small deviation of 3\% -- 5\% for below 1 PeV and $\sim$7\% at 3 PeV from the input spectra represented by the solid curves in Fig.~\ref{figure:systematic1}. If in reality the $H/He$ lies in the range covered by those model assumptions,  the real proton spectrum will be able to be reconstructed with an error smaller than 7\%. This actually turns out an estimate of the  systematic uncertainty due to the composition assumption, namely the \textcolor{black}{use} of the GSF model. In fact, the uncertainty is overestimated.
Once LHAASO measured the helium spectrum, the $H/He$ will be fixed. This uncertainty will be diminished through the iteration procedure.

\subsubsection{Purity of proton events}

\begin{figure}[htpb]
\centering\includegraphics[width=0.55\linewidth]{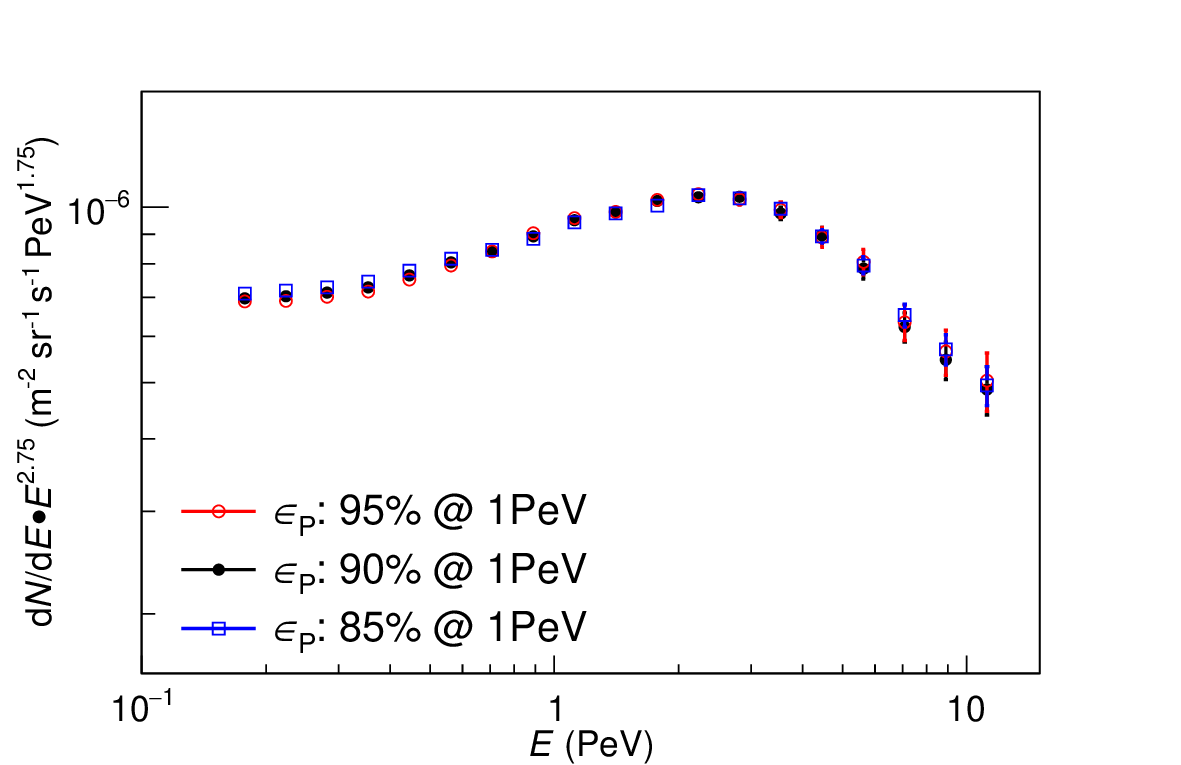}
\caption{By adjusting the proton selection criteria, samples with different proton event purities can be obtained. The black dots, red circles, and blue squares in the figure represent the energy spectrum results for proton event purities of 95\%, 90\%, and 85\% at 1 PeV, respectively. The flux difference among these spectra is less than 2.5\%.
}
\label{figure:systematic2}
\end{figure}

By implementing stricter criteria for proton sample selection, we can enhance purity of proton events but at the expense of reduced statistics. We adjust these standards in order to produce different proton event samples. The resulting purities are 95\%, 90\%, and 85\% at 1 PeV, corresponding to the selection efficiencies of 18\%, 25\%, and 35\%, respectively. Fig.~\ref{figure:systematic2} illustrates proton spectra corresponding to these selection efficiencies, revealing about 2.5\% spectral difference among the spectra.

\subsubsection{Observational environment}

We divided the data into two batches based on different scenarios. The events were categorized into two samples with similar event-count statistics according to atmospheric pressure: one sample with atmospheric pressure greater than 595 hPa and another with pressure less than 595 hPa. When comparing the energy spectra of these samples to those of the overall dataset, the observed difference is less than 1\%.

Additionally, we divided the data into two samples based on the intensity of sky background light, since moonlight is the dominant source of this background. One sample included events when the moon is above the horizon, while the other contains events with the moon below the horizon. The difference in energy spectra between these two samples and the overall dataset is found to be less than 2\%.

We categorized the data based on absolute humidity, creating two sets: one for absolute humidity greater than $1.53\times10^{-6}\mathrm{g/cm^{3}}$
and the other for humidity less than this value. The difference in energy spectra between these samples and the overall dataset is less than 1\%.

\subsection{Uncertainties in the shower energy measurement}
\subsubsection{Calibration}

The WFCTA employs a portable probe that has been absolutely calibrated by the National Institute of Metrology, China, to calibrate the light intensity of five different wavelengths of LEDs mounted inside the telescope~\cite{WFCTA-LED-calibration}.
This calibration ensures accurate calibration and effective monitoring of the camera.
After the initial calibration, the LEDs are used for absolute calibration and continuous monitoring of the camera's response.

Calibration errors can be categorized into two types. One type arises from the probe calibrated by the National Institute of Metrology, with an associated uncertainty of approximately 2.1\%. This type of error cannot be reduced through measurements from multiple telescopes. 

The other type of error results from factors such as probe placement, gain measurement, and the distribution of LED light sources, about 1.5\% in total.
Additionally, the measurement error related to the mirror reflectivity for each telescope is estimated to be around 1.2\%. The WFCTA consists of 18 telescopes that independently measure the energy spectrum of CRs. The total error caused by the transfer error from camera calibration and the measurement error of mirror reflectivity is $\sqrt{(1.5\%)^2 + (1.2\%)^2}/\sqrt{n-1}$$\approx $0.5\%, where $n$ represents the number of telescopes.

By substituting the  camera calibration errors into the energy reconstruction formula, the measurement error of energy is calculated to be approximately 1.5\%. Similarly, the energy measurement error caused by the measurement error of mirror reflectivity is about 1\%.
The errors caused by $N_\mu$ calibration, absolute humidity correction, aerosol correction, and air pressure correction are approximately 1\%, 1\%, 2\%, and 0.5\%, respectively.

\subsubsection{Hadronic interaction models}

There is an uncertainty related to the interaction models assumed in the simulation. By considering the high-energy hadronic interaction models QGSJETII-4, EPOS-LHC, and SIBYLL 2.3d, the uncertainty is found to be about 1.7\%.

\subsection{Summary of systematic uncertainties}

Table~\ref{table:systematicUncertainty} summarizes the different types of systematic uncertainties mentioned above. The total systematic uncertainty on the spectrum flux in Table~\ref{table:systematicUncertainty} is 5.8\% at 0.3 PeV and 8.5\% at 3 PeV, which is also the one shown in Table~\ref{table:protonSpectrumText}.
The systematic uncertainty introduced by the hadronic interaction models can be estimated through the differences in energy spectra obtained from three hadronic interaction models, with an uncertainty of 17\%.
The total systematic uncertainty on the energy measurement is about 4\%. 

\begin{table}[htbp]
\renewcommand{\arraystretch}{1.25}
\begin{center}
\begin{threeparttable}
\caption{Summary of systematic uncertainties.}
\begin{tabular}{lr}
\hline
\hline
1. Systematic uncertainties in the energy measurement:&\\
\hline
SiPM camera calibration & $\sim1.5\%$\\
Mirror reflectivity calibration & $\sim1\%$\\
$N_\mu$ calibration & $\sim1\%$\\
Absolute humidity & $\sim1\%$\\
Aerosol & $\sim2\%$\\
Air pressure & $\sim0.5\%$\\
Hadronic interaction models & $\sim1.7\%$\\
\hline
\hline
2. Systematic uncertainties in the spectrum flux&\\
\hline
Hadronic interaction models & $\sim$10\%@0.3 PeV, $\sim$15\%@3PeV$^1$\\
Composition models & $\sim$4\%@0.3 PeV, $\sim$7\%@3PeV$^2$\\
Purity of proton events & $\lesssim2.5\%$  \\
Air pressure & $\lesssim1\%$  \\
Background light & $\lesssim2\%$ \\
Absolute humidity & $\lesssim1\%$  \\
\hline
\end{tabular}
\begin{tablenotes}
    \item[1] The values are derived from  the maximum differences obtained when comparing QGSJETII-04 and SIBYLL 2.3d with EPOS-LHC, respectively.
    \item[2] The values are derived from  the maximum differences obtained when comparing Gaisser, Horandel and LVBI with GSF, respectively. 
\end{tablenotes}
\label{table:systematicUncertainty}
\end{threeparttable}
\end{center}
\end{table}

\section{Test of hadronic interaction models}
\begin{figure}[htpb]
\centering\includegraphics[width=0.98\linewidth]{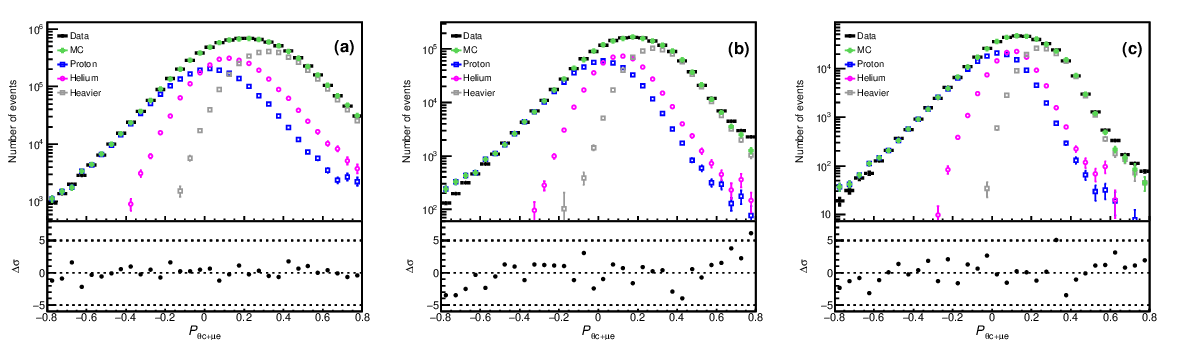}
\caption{Parameter comparison. These three sets of figures represent the comparisons of $P_{\mathrm{\theta c+\mu e}}$ in the energy ranges of $10^{-0.8}\sim10^{-0.4}$ PeV (a), $10^{-0.4}\sim10^{0.0}$ PeV (b), and $10^{0.0}\sim10^{1.1}$ PeV (c). The above row shows the distribution of $P_{\mathrm{\theta c+\mu e}}$ in both the data and the simulation in different energy intervals. The simulation events are based on EPOS-LHC model. The black dots represent the data, while the blue squares represent proton events, pink circles represent helium events, gray squares represent events heavier than helium, and green dots represent the sum of all simulated events. The row below shows the deviation between the data and the simulation.}
\label{figure:Parameter-comparison}
\end{figure}

\begin{figure}[htpb]
\centering\includegraphics[width=0.98\linewidth]{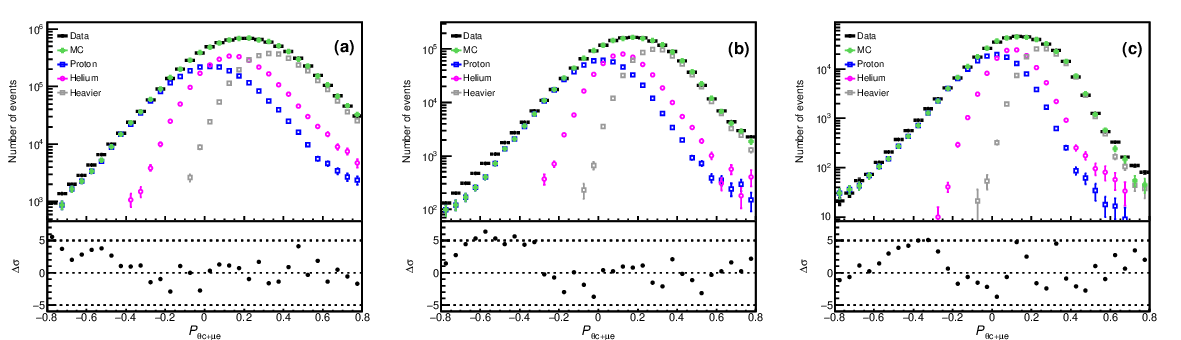}
\caption{Parameter comparison. These three sets of figures represent the comparisons of $P_{\mathrm{\theta c+\mu e}}$ in the energy ranges of $10^{-0.8}\sim10^{-0.4}$ PeV (a), $10^{-0.4}\sim10^{0.0}$ PeV (b), and $10^{0.0}\sim10^{1.1}$ PeV (c). The above row shows the distribution of $P_{\mathrm{\theta c+\mu e}}$ in both the data and the simulation in different energy intervals. The simulation events are based on QGSJETII-04 model. 
The black dots represent the data, while the blue squares represent proton events, pink circles represent helium events, gray squares represent events heavier than helium, and green dots represent the sum of all simulated events.
The row below shows the deviation between the data and the simulation.}
\label{figure:Parameter-comparisonQGSJET}
\end{figure}

\begin{figure}[htpb]
\centering\includegraphics[width=0.98\linewidth]{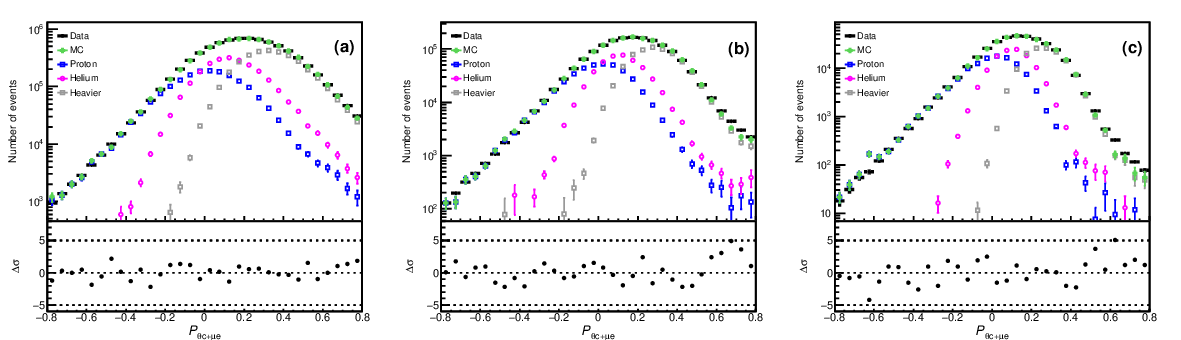}
\caption{Parameter comparison. These three sets of figures represent the comparisons of $P_{\mathrm{\theta c+\mu e}}$ in the energy ranges of $10^{-0.8}\sim10^{-0.4}$ PeV (a), $10^{-0.4}\sim10^{0.0}$ PeV (b), and $10^{0.0}\sim10^{1.1}$ PeV (c). The above row shows the distribution of $P_{\mathrm{\theta c+\mu e}}$ in both the data and the simulation in different energy intervals. The simulation events are based on SIBYLL 2.3d model. 
The black dots represent the data, while the blue squares represent proton events, pink circles represent helium events, gray squares represent events heavier than helium, and green dots represent the sum of all simulated events.
The row below shows the deviation between the data and the simulation.}
\label{figure:Parameter-comparisonSIBYLL}
\end{figure}

We conducted a comparison of the proton selection parameter $P_{\mathrm{\theta c+\mu e}}$ between the experimental data and the simulation data.
Fig.~\ref{figure:Parameter-comparison}, Fig.~\ref{figure:Parameter-comparisonQGSJET}, and Fig.~\ref{figure:Parameter-comparisonSIBYLL} show distributions of $P_{\mathrm{\theta c+\mu e}}$ for different types of events based on three hadronic models.
In the leftmost region, the distribution of $P_{\mathrm{\theta c+\mu e}}$ aligns closely between the experimental data and the proton events in the simulation.
As $P_{\mathrm{\theta c+\mu e}}$ increases, the influence of contamination from other components becomes apparent, leading to a higher distribution of $P_{\mathrm{\theta c+\mu e}}$ in the data compared to the proton events simulation. 

LHAASO can measure muons and parameters related to the $X_{\mathrm{max}}$ in an EAS. These parameters depend on the assumptions of the hadronic interaction models.
We can use parameters measured by LHAASO to test hadronic interaction models.
From Fig.~\ref{figure:Parameter-comparison} to Fig.~\ref{figure:Parameter-comparisonSIBYLL}, the distribution of $P_{\mathrm{\theta c+\mu e}}$ in the QGSJETII-04 model is less consistent with experimental observations than that of other models.
However, this analysis is not thorough enough to definitively rule out either model.
This paper focuses on the measurement of the proton energy spectrum and will not delve into the details of the study of hadronic interaction models.
Further investigation is necessary for LHAASO to contribute to models of hadronic interactions.



\begin{thebibliography}{99}
\bibitem{cosmicray-knee-1959}
Kulikov G, Khristiansen G. On the size spectrum of extensive air showers. J Exp Theor Phys 35 (1959) 635.
\bibitem{LHAASO-allparticle}
LHAASO Collaboration, Cao Z, et al. Measurements of all-particle energy spectrum and mean logarithmic mass of cosmic rays from 0.3 to 30 PeV with LHAASO-KM2A. Phys Rev Lett 132 (2024) 131002.

\bibitem{Gaisser:2013bla}
Gaisser TK, Stanev T, Tilav S. Cosmic ray energy spectrum from measurements of air showers. Front Phys 8 (2013) 748–758.
\bibitem{Kachelriess-2019}
Kachelrieß M, Semikoz M. Cosmic ray models. Prog Part Nucl Phys 109 (2019) 103710.
\bibitem{LHAASO-catalogue}
LHAASO Collaboration, Cao Z, et al. The first LHAASO catalog of gamma-ray sources. Astrophys J Suppl Ser 271 (2024) 25.
\bibitem{LHAASO-nature-12}
LHAASO Collaboration, Cao Z, et al. Ultrahigh-energy photons up to 1.4 petaelectronvolts from 12 $\gamma$-ray galactic sources. Nature 594 (2021) 33–36.
\bibitem{LHAASO-sygnus}
LHAASO Collaboration, Cao Z, et al. An ultrahigh-energy $\gamma$-ray bubble powered by a super PeVatron. Sci Bull 69 (2024) 449–457.
\bibitem{O.AdrianiandL.Pacini2023}
Adriani O, Pacini L. Results from high energy direct measurements and future prospects. EPJ Web Conf 283 (2023) 02001.
\bibitem{DAMPE:2019gys}
DAMPE Collaboration, An Q, et al. Measurement of the cosmic ray proton spectrum from 40 GeV to 100 TeV with the DAMPE satellite. Sci Adv 5 (2019) eaax3793.
\bibitem{CALET:2022vro}
CALET Collaboration, Adriani O, et al. Observation of spectral structures in the flux of cosmic-ray protons from 50 GeV to 60 TeV with the Calorimetric Electron Telescope on the international space station. Phys Rev Lett 129 (2022) 101102.
\bibitem{Choi:2022aht}
ISS-CREAM Collaboration, Choi GH, et al. Measurement of high-energy cosmic-Ray proton spectrum from the ISS-CREAM experiment. Astrophys J 940 (2022) 107.
\bibitem{CASAMIA-nima1994}
Borione A, Covault CE, Cronin JW, et al. A large air shower array to search for astrophysical sources emitting $\gamma$-rays with energies $\geq$ 10$^{14}$ eV. Nucl Instrum Methods Phys Res A 346 (1994) 329–352.
\bibitem{GRAPES-3:2024mhy}
GRAPES-3 Collaboration, Varsi F, et al. Evidence of a hardening in the cosmic ray proton spectrum at around 166 TeV observed by the GRAPES-3 experiment. Phys Rev Lett 132 (2024) 051002.
\bibitem{kaskade-spectrum-00}
KASCADE Collaboration, Antoni T, et al. KASCADE measurements of energy spectra for elemental groups of cosmic rays: Results and open problems. Astropart Phys 24 (2005) 1–25.
\bibitem{kaskade-spectrum-01}
KASCADE Collaboration, Apel WD, et al. Energy spectra of elemental groups of cosmic rays: Update on the KASCADE unfolding analysis. Astropart Phys 31 (2009) 86–91.
\bibitem{kaskade-spectrum-02}
KASCADE-Grande Collaboration, Apel WD, et al. KASCADE-Grande measurements of energy spectra for elemental groups of cosmic rays. Astropart Phys 47 (2013) 54–66.
\bibitem{icetop-spectrum}
IceCube Collaboration, Aartsen MG, et al. Cosmic ray spectrum and composition from PeV to EeV using 3 years of data from IceTop and IceCube. Phys Rev D 100 (2019) 082002.
\bibitem{LHAASO-Design}
LHAASO Collaboration, Cao Z, et al. A future project at tibet: the large high altitude air shower observatory (LHAASO). Chin Phys C 34 (2010) 249.
\bibitem{LHAASO-detectors}
LHAASO Collaboration, He HH, et al. Design of the LHAASO detectors. Radiat Detect Technol Methods 2 (2018) 7.
\bibitem{LHAASO-science-book2}
LHAASO Collaboration, Piazzoli BD, et al. Chapter 4 cosmic-ray physics. Chin Phys C 46 (2022) 030004.
\bibitem{LHAASO-science-book1}
LHAASO Collaboration, Ma XH, et al. Chapter 1 LHAASO instruments and detector technology. Chin Phys C 46 (2022) 030001.
\bibitem{LHAASO-XmaxRec}
LHAASO Collaboration, You ZY, et al. The Energy spectrum of cosmic ray proton and helium above 100 TeV measured by LHAASO experiment. PoS ICRC2021 (2021) 377.
\bibitem{EASmodel-Matthews}
Matthews J. A heitler model of extensive air showers. Astropart Phys 22 (2005) 387–397.
\bibitem{KM2A-ED-calibration}
LHAASO Collaboration, Aharonian F, et al. Self-calibration of LHAASO-KM2A electromagnetic particle detectors using single particles within extensive air showers. Phys Rev D 106 (2022) 122004.
\bibitem{KM2A-MD-calibration}
LHAASO Collaboration, Zuo X, et al. Design and performances of prototype muon detectors of LHAASO-KM2A. Nucl Instrum Methods Phys Res A 789 (2015) 143–149.
\bibitem{WFCTA-LED-calibration}
LHAASO Collaboration, Aharonian F, et al. Absolute calibration of LHAASO WFCTA camera based on led. Nucl Instrum Methods Phys Res A 1021 (2022) 165824.
\bibitem{QGSJETII-04}
Ostapchenko S. QGSJET-II: physics, recent improvements, and results for air showers. EPJ Web Conf 52 (2013) 02001.
\bibitem{EPOS-LHC}
Pierog T, Karpenko I, Katzy JM, et al. EPOS LHC: Test of collective hadronization with data measured at the cern large hadron collider. Phys Rev C 92 (2015) 034906.
\bibitem{SIBYLL2.3d}
Riehn F, Engel R, Fedynitch A, et al. Hadronic interaction model sibyll 2.3d and extensive air showers. Phys Rev D 102 (2020) 063002.
\bibitem{AMS2021-spectrum}
AMS Collaboration, Aguilar M, et al. The alpha magnetic spectrometer (AMS) on the international space station: Part II - results from the first seven years. Phys Rep 894 (2021) 1–116.

\bibitem{Atkin:2018wsp}
NUCLEON Collaboration, Atkin E, et al. New universal cosmic-ray knee near a magnetic rigidity of 10 TV with the NUCLEON space observatory. JETP Lett 108 (2018) 5–12.
\bibitem{Liu:2018fjy}
Liu W, Guo YQ, Yuan Q. Indication of nearby source signatures of cosmic rays from energy spectra and anisotropies. J Cosmol Astro Part Phys 2019 (2019) 010.
\bibitem{Fang:2020cru}
Fang K, Bi XJ, Yin PF. DAMPE proton spectrum indicates a slow-diffusion zone in the nearby ISM. Astrophys J 903 (2020) 69.
\bibitem{Li:2021szb}
Li AF, Yuan Q, Liu W, et al. Large-scale anisotropy of galactic cosmic rays as a probe of local cosmic-ray propagation. Astrophys J 962 (2024) 43.
\bibitem{Qiao:2022cge}
Qiao BQ, Luo Q, Yuan Q, et al. Understanding the phase reversals of galactic cosmic-ray anisotropies. Astrophys J 942 (2022) 13.
\bibitem{Zhang:2022pzt}
Zhang YR, Liu SM, Zeng HD. A three-component model for cosmic ray spectrum and dipole anisotropy. Mon Not R Astron Soc 511 (2022) 6218–6224.
\bibitem{Prevotat:2025ktr}
Prevotat C, Zhu Z, Koldobskiy S, et al. Diffuse gamma-ray and neutrino emission from the milky way and the local knee in the cosmic ray spectrum. arXiv: 2507.10823 (2025).
\bibitem{Zhang:2025tew}
Zhang BT, Kimura SS, Murase K. Microquasar jet-cocoon systems as pevatrons. arXiv: 2506.20193 (2025).
\bibitem{Bell:2013kq}
Bell AR, Schure KM, Reville B, et al. Cosmic-ray acceleration and escape from supernova remnants. Mon Not R Astron Soc 431 (2013) 415–429.
\bibitem{Sveshnikova:2003sa}
Sveshnikova LG. The knee in the galactic cosmic ray spectrum and variety in supernovae. Astron Astrophys 409 (2003) 799–807.
\bibitem{Strong:2010pr}
Strong AW, Porter TA, Digel SW, et al. Global cosmic-ray-related luminosity and energy budget of the milky way. Astrophys J Lett 722 (2010) L58.
\bibitem{LHAASO:2024psv}
LHAASO Collaboration, Cao Z, et al. Ultrahigh-energy gamma-ray emission associated with black hole-jet systems. arXiv: 2410.08988 (2024).
\bibitem{Alfaro:2024cjd}
HAWC Collaboration, Alfaro R, et al. Ultra-high-energy gamma-ray bubble around microquasar v4641 sgr. Nature 634 (2024) 557–560.
\bibitem{Wang:2025yqy}
Wang JS, Reville B, Aharonian F. Galactic superaccreting x-ray binaries as super-pevatron accelerators. Astrophys J Lett 989 (2025) L25.


\bibitem{KM2A-monitor}
LHAASO Collaboration, Cao Z, et al. Data quality control system and long-term performance monitor of LHAASO-KM2A. Astropart Phys 164 (2025) 103029.
\bibitem{KM2AcrabCPC}
LHAASO Collaboration, Aharonian F, et al. Observation of the crab nebula with LHAASO-KM2A a performance study. Chin Phys C 45 (2021) 025002.
\bibitem{CORSIKA-PACKAGE}
Heck D, Knapp J, Capdevielle JN, et al. CORSIKA: A Monte Carlo code to simulate extensive air showers, 6019, Forschungszentrum Karlsruhe GmbH, Karlsruhe, FZKA, 1998
\bibitem{KM2A-Geant4}
LHAASO Collaboration, Cao Z, et al. LHAASO-KM2A detector simulation using Geant4. Radiat Detect Technol Methods 8 (2024) 1437–1447.
\bibitem{GEANT4}
Geant4 Collaboration, Agostinelli S, et al. Geant4—a simulation toolkit. Nucl Instrum Methods Phys Res A 506 (2003) 250–303.
\bibitem{fluka}
FLUKA collaboration, Battistoni G, et al. Overview of the fluka code. Ann Nucl Energy 82 (2015) 10–18.
\bibitem{WFCTA-construction}
LHAASO Collaboration, Aharonian F, et al. Construction and on-site performance of the LHAASO WFCTA camera. Eur Phys J C 81 (2021).
\bibitem{WLP-Erec}
Wang LP, Ma LL, Zhang SS, et al. Cosmic ray mass independent energy reconstruction method using Cherenkov light and muon content in LHAASO. Phys Rev D 107 (2023) 043036.

\bibitem{EASmodel}
Hörandel JR. Cosmic rays from the knee to the second knee: $10^{14}$ to $10^{18}$eV. Mod Phys Lett A 22 (2007) 1533–1551.

\bibitem{GSFmodel}
Dembinski H, Engel R, Fedynitch A, et al. Data-driven model of the cosmic-ray flux and mass composition from 10 GeV to $10^{11}$ GeV. PoS ICRC2017 (2017) 533.

\bibitem{Horandelmodel}
Hörandel JR. On the knee in the energy spectrum of cosmic rays. Astropart Phys 19 (2003) 193–220.
\bibitem{BXJmodel}
Lv XJ, Bi XJ, Fang K, et al. Precise measurement of the cosmic-ray spectrum and ⟨ln$A$⟩ by lhaaso -- connecting the galactic to the extragalactic components. arXiv: 2403.11832 (2024).

\end{thebibliography}
\end{document}